\documentclass[onecolumn,amsmath,amssymb,superscriptaddress]{revtex4}
\usepackage{graphicx}
\usepackage{amssymb}
\usepackage{epstopdf}
\usepackage{bbold}
\usepackage{xcolor}

\begin{document}
\title{Mechanics of active surfaces}
\author{Guillaume Salbreux}
\affiliation{Max Planck Institute for the Physics of Complex Systems, N\"othnitzer Str. 38, 01187 Dresden, Germany}
\affiliation{The Francis Crick Institute, 1 Midland Road, NW1 1AT, United Kingdom}
\author{Frank J\"ulicher}
\affiliation{Max Planck Institute for the Physics of Complex Systems, N\"othnitzer Str. 38, 01187 Dresden, Germany}
\date{\today}

\begin{abstract}
We derive a fully covariant theory of the mechanics of active surfaces. 
This theory provides a framework for the study of active
biological or chemical processes at surfaces, such as the cell cortex, the
mechanics of epithelial tissues, or reconstituted active systems on surfaces.
We introduce forces and torques acting on a surface, and derive the associated 
force balance conditions. 
We show
that surfaces with in-plane rotational symmetry 
can have broken up-down, chiral or planar-chiral symmetry. 
We discuss the rate of entropy production
in the surface and 
write linear constitutive relations that satisfy the Onsager relations. 
We show that the bending modulus, the spontaneous curvature and 
the surface tension of a passive surface are renormalised by 
active terms. Finally, we identify novel active terms which are not found in a passive theory 
and discuss examples of shape instabilities that are related to active processes in the surface.
\end{abstract}
\maketitle

\begin{figure}[h!]
 \centering
 \includegraphics[width=10cm]{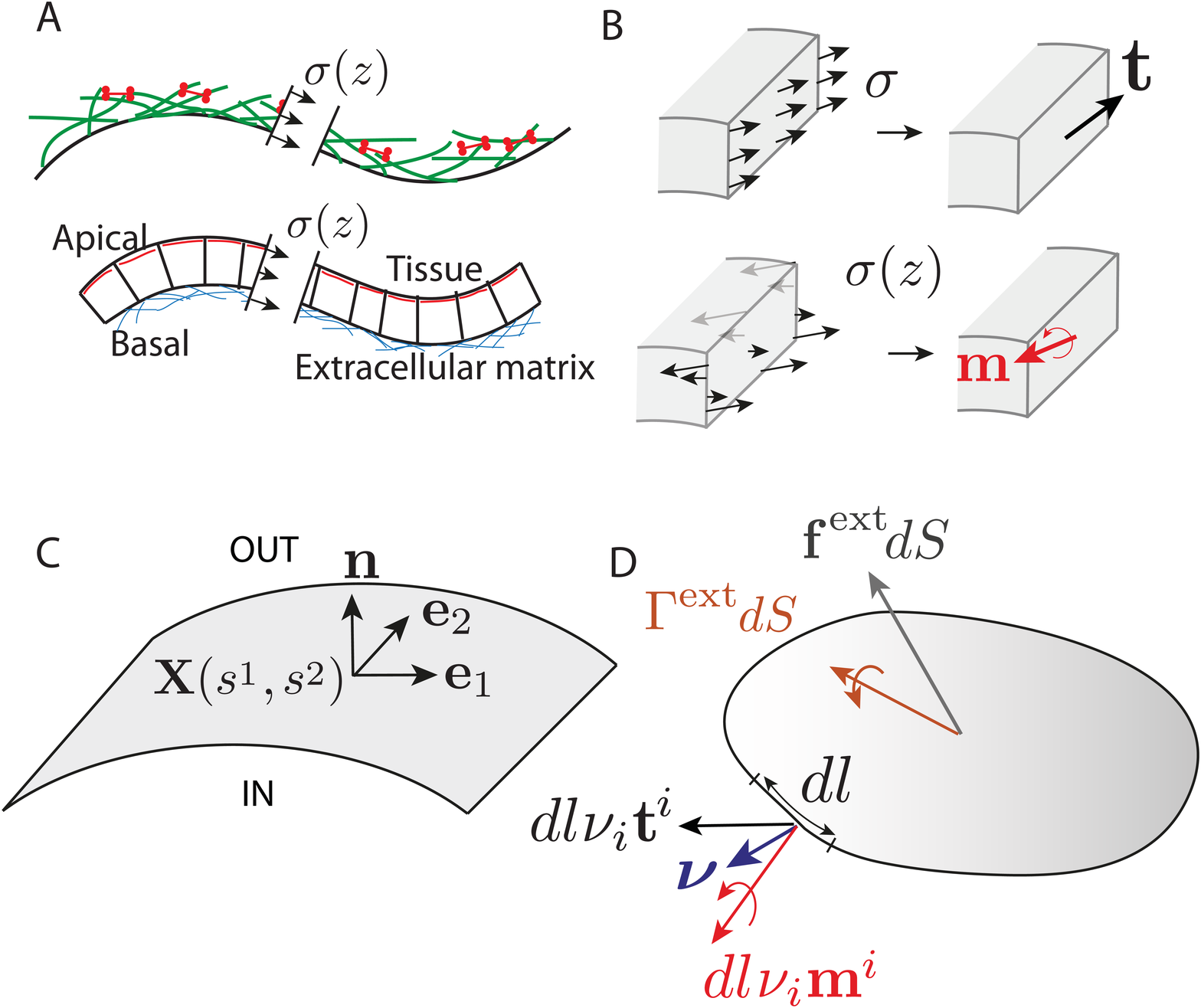}
 \caption{
 \label{fig:schematic}A. Filaments and motors near a surface and epithelial tissues are examples of active surfaces. B. The distribution of stresses within a thin layer give rise to stresses and torques when integrated across the thickness of the layer. C. Local basis of tangent vectors $\mathbf{e}_1$, $\mathbf{e}_2$ and normal vector $\mathbf{n}$ associated to the surface $\mathbf{X}(s^1,s^2)$. D. Internal and external forces and torques acting on a surface element with surface area $dS$.
 } 
\end{figure}

Biological systems exhibit a stunning variety of complex morphologies and shapes.
Organisms form from a fertilized egg in a dynamic process called morphogenesis.
Such shape forming processes in biology involve active mechanical events during which surfaces
undergo shape changes that are driven by active stresses and torques generated in
the material. Important examples are two-dimensional tissues, so called epithelia. They
represent surfaces that can deform their shape as a result of active cellular processes \cite{lecuit2007cell}. 
Cells also exhibit a variety of different shapes and can undergo active
shape changes. For example during cell division, cells round up to a spherical shape
due to an increase of active surface tension \cite{kunda2009actin}. Cell shapes are governed by the cell cortex,
a thin layer of an active contractile material at the surface of the cell \cite{salbreux2012actin}. Epithelial tissues and the cell surface are
examples of active surfaces. In addition, recent experiments have reconstituted thin shells of active material in-vitro \cite{keber2014topology}.
These are thin sheets of active matter that can deform due to
the generation of internal forces and torques that are balanced by external forces (Figure \ref{fig:schematic}A). 

The theory of active gels describe the large-scale properties of viscoelastic matter driven
out-of-equilibrium due to a source of chemical free energy in the system \cite{kruse2005generic}. 
A number of processes in living systems have been successfully described using this theoretical framework \cite{prost2015active}. Living or artificial active systems often assemble into nearly 
two-dimensional surfaces. 
 To understand the physics of such active surfaces, requires a systematic analysis of force and torque balances in curved two-dimensional 
geometries, taking into account active stresses and material properties.
The shapes of passive fluid membranes has been described with considerable success 
by the Helfrich free energy, a coarse-grained description of membranes 
with an expansion of the free energy in powers of the curvature tensor \cite{helfrich1973elastic}. 
Expressions for the stress and torque tensors within a Helfrich membrane have been obtained.  
The associated force and torque balance equations are equivalent to shape equations for miminal
energy shapes \cite{capovilla2002stresses, fournier2007stress}. Active membranes theories have expanded the description of passive membranes to include external forces induced by pumps contained in a membrane \cite{ramaswamy2000nonequilibrium,chen2004internal,gov2004membrane}.


The morphogenesis of epithelial tissues is a highly complex problem involving forces generated actively within the cells. Distribution of forces acting along the cross-section of a sheet-like tissue give rise to in-plane tensions, but also to internal torques resulting from differential stresses acting along the cross-section of the tissue (Fig. \ref{fig:schematic}B). These differential stresses are crucial to generate tissue shape changes \cite{sawyer2010apical}. However, no framework currently allows to describe the mechanics of active thin surfaces with internal stresses and torque densities.

In this work, we present such a general framework for the mechanics of active surfaces, driven internally out of equilibrium by 
molecular processes such as a chemical reaction. We start by considering forces and torques generated in a surface of arbitrary shape. The corresponding expression for the virtual work shows that components of the tension and torque tensors are coupled to the variation of the metric, of the 
curvature tensors and of the  Christoffel symbols 
defined for the surface (Eq. \ref{VirtualWork2}). Using these expressions, 
we then derive the entropy production for a fluid surface undergoing chemical reactions. 
We analyse the symmetries of surfaces with rotational symmetry 
in the plane, and show that they can have up-down, chiral or planar-chiral 
broken symmetry. We write down the corresponding constitutive equations 
for the components of the tension and torque tensors 
and for the fluxes of the chemical species. Interestingly, the generic 
constitutive equations involve couplings of the curvature tensor 
with the chemical potential of the surface chemical species. We then 
discuss the stability of a flat active fluid with broken up-down symmetry. Finally, we show that generic equations for an active elastic thin shell can be obtained using the same framework.

\section{Force and torque balance on a curved surface}

We consider a curved surface $\mathbf{X}(s^1,s^2)$ parametrised by two generalised coordinates $s^1$, $s^2$ (Figure \ref{fig:schematic}C). We use latin indices to refer to surface coordinates and greek indices to refer to 3D euclidean coordinates. We introduce the metric tensor $g_{ij}=\mathbf{e}_i\cdot\mathbf{e}_j$ where $\mathbf{e}_i=\partial_i \mathbf{X}$ with $\partial_i=\partial/\partial s^i$. The curvature tensor is defined as $C_{ij}=-\partial_i \partial_j \mathbf{X} \cdot\mathbf{n}$, where $\mathbf{n}=\mathbf{e}_1\times\mathbf{e}_2/|\mathbf{e}_1\times\mathbf{e}_2|$ is the unit normal vector, which we usually consider to point outward for a closed surface.

We denote $dl$ with $dl^2=g_{ij}ds^i ds^j$ a line element on the surface, and $dS=\sqrt{g} ds^1 ds^2$  a surface element, where $g=\det g_{ij}$ is the determinant of the metric tensor (Appendix \ref{AppendixDifferentialGeometry}).


The force $\textbf{f}$ and torque $\boldsymbol{\Gamma}$ across a line of length $dl$ with {\it unit} vector $\boldsymbol{\nu}=\nu^i \mathbf{e}_i$, tangential to the surface and normal to the line 
can be expressed as
\begin{align}
\label{DefinitionsTensionMoment}
\mathbf{f}&=dl\; \nu^i \textbf{t}_i=dl\; \nu_i \textbf{t}^i,\\
\boldsymbol{\Gamma}&=dl\; \nu^i \textbf{m}_i=dl\; \nu_i \textbf{m}^i \quad ,
\end{align}
where we have introduced the tension ${\bf t}_i$ and moment ${\bf m}_i$ per unit length (Figure \ref{fig:schematic}B,D). Decomposing $\textbf{t}^i$ and $\textbf{m}^i$  in tangential and normal components as 
\begin{align}
\textbf{t}^i&=t^{ij} \textbf{e}_j+t_{n}^i \textbf{n},\label{TensionProjection}\\
\textbf{m}^i&=m^{ij}\textbf{e}_j+m^i_{n} \mathbf{n} \quad , \label{MomentProjection}
\end{align}
defines the tension and moment per unit length tensors $t^{ij}$, $t_n^i$, $m^{ij}$ and $m_n^i$.

By expressing the total force acting on a region of surface $\mathcal{S}$ with contour $\mathcal{C}$ and using Newton's law, one finds
\begin{equation}
\int_\mathcal{S} dS \rho \mathbf{a}=\oint_\mathcal{C} dl \nu_i \mathbf{t}^i+\int_\mathcal{S} dS \mathbf{f}^{\rm{\rm{ext}}},
\end{equation}
where $\rho$ is the surface mass density, $\mathbf{a}$ is the local center-of-mass acceleration, $\mathbf{f}^{\rm{ext}}$ is an external force surface density. When the surface is embedded in a medium, the external force surface density is related to stresses exerted by the medium on the surface, $f^{\rm{ext}}_{\alpha}=\sigma_{\beta\alpha}n_{\beta}$ with $\sigma_{\alpha\beta}$ the 3-dimensional stress tensor in the medium.
The total torque obeys
\begin{align}
\int_\mathcal{S} dS[\mathbf{X}\times \rho \mathbf{a}]&=\oint_\mathcal{C} dl \nu_i \left[\mathbf{m}^i+\mathbf{X}\times\mathbf{t}^i\right]+\int_\mathcal{S} dS \left[\boldsymbol{\Gamma}^{\rm{ext}}+\mathbf{X}\times\mathbf{f}^{\rm{ext}}\right].
\end{align}
where $\boldsymbol{\Gamma}^{\rm{ext}}$ is the external torque surface density, and where the left-hand side is the torque stemming from inertial forces. Here, we ignore the moment of inertia tensor for simplicity. 
This results in the force balance expression (Appendix \ref{ForceBalanceDerivation}):
\begin{align}
\nabla_i \mathbf{t}^i&=-\mathbf{f}^{\rm{ext}}+\rho \mathbf{a}
\label{ForceBalanceEquation},\\
\nabla_i \mathbf{m}^i&=\mathbf{t}^i\times\mathbf{e}_i-\boldsymbol{\Gamma}^{\rm{\rm{ext}}}\label{TorqueBalanceEquation}.
\end{align}
These equations can be expressed in terms of the components of the tension and torque tensors:
\begin{align}
\nabla_i t^{ij}+C_{i}{}^j t_{n}^i&=-f^{{\rm ext},j}+\rho a^j\label{ForceBalanceTangential},\\
\nabla_i t_{n}^i-C_{ij} t^{ij}&=-f^{\rm{ext}}_{n}+\rho a_{n}\label{ForceBalanceNormal},\\
\nabla_i m^{ij}+C_{i}{}^j m_{n}^i&=\epsilon_{i}{}^{j}   t_{n}^i-\Gamma^{{\rm ext},j} \label{TorqueBalanceTangential},\\
\nabla_i m_{n}^i-C_{ij} m^{ij}&=-\epsilon_{ij}t^{ij}-\Gamma^{\rm{ext}}_{n}\label{TorqueBalanceNormal},
\end{align}
where the tangential and normal component of a vector $\mathbf{v}$ on the surface are written $v^{i}=\mathbf{v}\cdot \mathbf{e}^i$ and $v_{n}=\mathbf{v}\cdot\mathbf{n}$. 

\section{Virtual work}

We introduce the virtual work $\delta W$, which is the mechanical work acting on a region of surface $\mathcal{S}$ enclosed by a contour $\mathcal{C}$, upon a small deformation $\delta \mathbf{X}$ of the surface, with $\mathbf{X'}(s^1,s^2)=\mathbf{X}(s^1,s^2)+\delta\mathbf{X}(s^1,s^2)$. Here $\delta \mathbf{X}(s^1,s^2)$ represents a displacement of a material point on the surface specified by $(s^1,s^2)$. The virtual work can be defined as
\begin{align}
\label{VirtualWorkDefinition}
\delta W=\oint_\mathcal{C} dl \mathbf{\nu}_i (\mathbf{t}^i\cdot \mathbf{\delta X}+\frac{1}{2} \boldsymbol{m}^{i}\cdot(\boldsymbol{\nabla}\times \delta\mathbf{X}))+\int_\mathcal{S} dS( (\mathbf{f}^{\rm{ext}}-\rho \mathbf{a})\cdot\delta\mathbf{X}
+\frac{1}{2} \boldsymbol{\Gamma}^{\rm{ext}} \cdot(\boldsymbol{\nabla}\times \delta\mathbf{X})),
\end{align}
where $\mathcal{S}$ is the surface region enclosed by $\mathcal{C}$, and we have introduced the rotational operator in euclidian space (Eq. \ref{AppendixCurlExpression}):
\begin{equation}
\label{CurlExpression}
(\boldsymbol{\nabla}\times\delta\mathbf{X})_{\alpha}= \epsilon_{\alpha\beta\gamma} e^i_{\beta}(\partial_{i} \delta X_{\gamma}) +\epsilon_{\alpha\beta\gamma} n_{\beta}(\partial_n \delta X_{\gamma}).
\end{equation} 
In Eq.\ref{CurlExpression}, we have introduced the normal derivative of the surface deformation, $\partial_n \delta \mathbf{X}$. We consider here $\partial_n \delta \mathbf{X}=-(\partial_i \delta \mathbf{X}\cdot\mathbf{n}) \mathbf{e}^i$ (Appendix \ref{AppendixVariationSurfaceQuantities}). 

The terms in the expression of the virtual work \ref{VirtualWorkDefinition} describe the work due to forces and torques acting at the boundary $\mathcal{C}$ as well as external forces and torques acting on the surface $\mathcal{S}$.
Using force balance and the divergence theorem, the virtual work can be re-expressed as (see Appendix \ref{DifferentialWork})
\begin{align}
\delta W
\label{VirtualWork2}
&=\int_\mathcal{S} dS\bigg(\overline{t}^{ij}\frac{\delta g_{ij}}{2}  + \overline{m}^{i}{}_{j}
\delta C_{i}{}^{j}+m^{i}_{n} \frac{\epsilon^{j}{}_{k} \delta \Gamma_{ij}^k}{2}   \bigg).
\end{align}
Here, the explicit expression of the metric variation $\delta g_{ij}$, curvature variation $\delta C_{i}{}^{j}$ and variation of Christoffel symbols $\delta \Gamma_{ij}^k$ as a function of the surface variation $\delta \mathbf{X}$ are given in Appendix \ref{AppendixVariationSurfaceQuantities}. We have introduced  the in-plane tension and bending moment tensors:
\begin{align}
\overline{t}^{ij}&=t^{ij}_s+\frac{1}{2}\left(\bar{m}^{ki}C_k{}^j+\bar{m}^{kj}C_k{}^i\right),\label{DefinitionInPlaneStressTensor}\\
\overline{m}^{ij}&=-m^{ik}\epsilon_{k}{}^{j}.\label{DefinitionInPlaneTorqueTensor}
\end{align}
where the $s$ subscript denotes the symmetric part of the tensor (Eq. \ref{TensorSymmetricAntisymmetricDecomposition}). In Eq. \ref{VirtualWork2}, we have used a reference frame that deforms with the material.

The virtual work given by equation \ref{VirtualWork2} can be interpreted physically as the mechanical work due to different types of deformations. The in-plane surface stress tensor $\bar{t}^{ij}$ is conjugate to the variation of the metric tensor $\delta g_{ij}$, describing internal shear and area compression. The in-plane tension tensor $\overline{t}_{ij}$ introduced in Eq. \ref{DefinitionInPlaneStressTensor} differs from the tension tensor $t_{ij}$ introduced in Eq. \ref{TensionProjection}: this is because in a thin shell, a deformation leading to a change of metric of the surface mid-plane corresponds to a three-dimensional shear within the shell. As a result, the work to deform the surface mid-plane depend on the in-plane bending moment tensor, which reflects the distribution of stresses across the thickness of the shell. The in-plane tensor $\bar{m}^{ij}$ of bending moments is conjugate to the variation of the curvature tensor $\delta C_{i}{}^{j}$ due to bending of the surface. The normal torque $m_n^i$ is conjugate to gradients of local rotations $\epsilon^{i}{}_{k}\delta\Gamma_{ij}^k$.
The expression of the virtual work \ref{VirtualWork2} does not include shear perpendicular to the surface: this would require the introduction of an additional variable.

The virtual work given in Eq. \ref{VirtualWork2} is very general. In order to evaluate the virtual work for a given surface deformation, the values of the internal stresses characterised by the in-plane stress tensor $\overline{t}^{ij}$, the in-plane bending moment tensor $\overline{m}^{ij}$, and the normal torque $\mathbf{m}_n$, have to be known. In general, they are provided by constitutive relations describing the properties of the material associated with the surface.

We now discuss constitutive relations for active fluid and elastic curved surfaces. The case of a passive membrane is discussed in Appendix \ref{PassiveMembrane}.

\section{Curved active film}

We now use concepts for irreversible thermodynamics to derive constitutive equations for a curved isotropic fluid. We consider a fluid consisting of several species $\alpha=1..N$ with concentrations $c^{\alpha}$.  The local mass density is given by $\rho=\sum_{\alpha} m^{\alpha} c^{\alpha}$ with $m^{\alpha}$ the molecular mass of species $\alpha$. The free energy density in the rest frame is denoted $f_0(c^{\alpha},C_{i}{}^{j},T)$ where $C_{i}{}^{j}$ is the curvature tensor of the film in mixed coordinates, and $T$ the temperature. The differential of $f_0$ is
\begin{equation}
\label{FluidMembraneFreeEnergyDensity}
df_0=\mu^{\alpha} d c^{\alpha}+K^{i}{}_{j} dC_{i}{}^{j}-sd T,\\
\end{equation}
where $\mu^{\alpha}$ is the chemical potential of component $\alpha$, $K^{i}{}_{j}$ is the passive bending moment and $s$ the entropy density.
The total free energy density is
\begin{equation}
\label{TotalFreeEnergyDensity}
f=\frac{1}{2}\rho v^2 + f_0,
\end{equation}
where the kinetic energy is given by $\frac{1}{2}\rho v^2=\frac{1}{2}\rho \left[v_i v^i+(v_n)^2\right] $. We denote $\mu_{tot}^{\alpha}=df/dc^{\alpha}=\mu^{\alpha}+m^{\alpha} v^2/2$ the total chemical potential of the chemical species $\alpha$.

\subsection{Conservation equations}

\begin{figure}[h!]
 \centering
 \includegraphics[width=6cm]{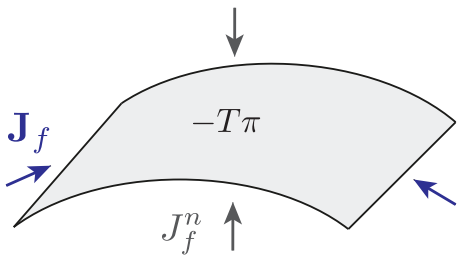}
 \caption{
 \label{fig:freeenergyfluxes} Free energy balance on a surface element in the isothermal case. Free energy density is exchanged between the surface element and the surrounding surface with a flux $\mathbf{J}^f$, with the surrounding bulk with flux $J_n^f$, and is produced with rate $- T \pi$.
 } 
\end{figure}

We start by deriving conservation equations for the surface mass, concentration of chemical species, energy, entropy and free energy. Using an Eulerian representation (Appendix \ref{EulerianLagrangian}), mass balance reads
\begin{eqnarray}
\label{MassBalanceEquation}
\partial_t \rho+\nabla_i(\rho v^i ) +  v_n C_i{}^i \rho=J_{n}^{\rho},
\end{eqnarray}
where $J_{n}^{\rho}$ is a source term due to mass exchange with the environment and $\mathbf{v}=v^i \mathbf{e}_i+v_n \mathbf{n}$ is the center-of-mass velocity.

The concentrations $c^{\alpha}$ obey the balance equation
\begin{eqnarray}
\label{ConcentrationBalanceEquation}
\partial_t c^{\alpha}+\nabla_i J^{\alpha i}+v_n C_i{}^i c^{\alpha}=J^{\alpha}_n+r^{\alpha},
\end{eqnarray}
where $J^{\alpha i}=c^{\alpha} v^i+{j^{\alpha}}^i$ is the tangential flux in the surface of molecule $\alpha$, $j^{\alpha,i}$ is the flux relative to the center of mass, $J^{\alpha}_n$ describes exchanges between the surface and its surrounding environment, and $r^{\alpha}$ denote source and sink terms corresponding to chemical reactions in the surface. Mass conservation implies the following relation between fluxes of molecules and chemical rates
\begin{align}
\sum_{\alpha} m^{\alpha} J_n^{\alpha}&=J_n^{\rho},\label{BalanceNormalFluxes}\\
\sum_{\alpha} m^{\alpha} j^{\alpha,i}&=0,\label{BalanceTangentialFluxes}\\
\sum_{\alpha} m^{\alpha} r^{\alpha}&=0.\label{BalanceChemicalReactions}
\end{align}
In the remaining of this work, summation over $\alpha$ is implicit. The conservation of energy and the balance of entropy and free energy density have the form (Figure \ref{fig:freeenergyfluxes})
\begin{eqnarray}
\partial_t e+\nabla_i J^{e,i}+v_n C_i{}^i e &=&J^e_n\label{BalanceInternalEnergy},\\
\partial_t s+\nabla_i J^{s,i}+v_n C_i{}^i s &=&J^s_n+\pi\label{BalanceEntropy},\\
\partial_t f+\nabla_i J^{f,i}+v_n C_i{}^i f&=&J^f_n-T\pi-J_s^i\nabla_i T-\partial_t T s\label{BalanceFreeEnergy},
\end{eqnarray}
where $e$ and $s$ are the energy and entropy density respectively, $J^e_n$ and $J^s_n$ are energy and entropy fluxes entering the surface from the adjacent bulk,  $J^{e,i}$ and $J^{s,i}$ are tangential energy and entropy fluxes within the surface, and $J^f_n=J^e_n-TJ^s_n$ and $J^{f,i}=J^{e,i}-TJ^{s,i}$ are the normal and tangential fluxes of free energy. The entropy production rate within the surface is denoted $\pi$. Eq. \ref{BalanceFreeEnergy} is obtained from the relation $f=e-Ts$ and Eqs. \ref{BalanceInternalEnergy} and \ref{BalanceEntropy}. In the following, we consider for simplicity the isothermal case.


\subsection{Translation and rotation invariance}

We now discuss relations between equilibrium tensions and torques implied by invariance of the surface properties under a rigid translation or rotation.

\subsubsection{Gibbs-Duhem relation}

\begin{figure}[h!]
 \centering
  \includegraphics[width=10cm]{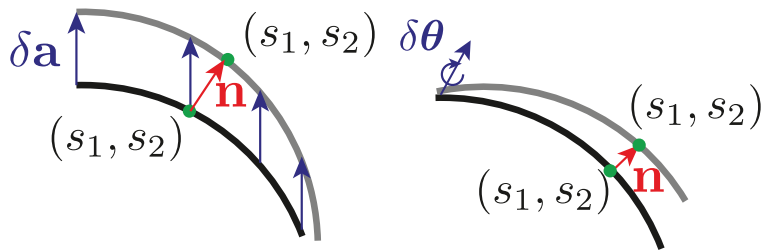}
 \caption{
 \label{fig:schematicGibbsDuhem}Two Gibbs-Duhem relations for a fluid surface are obtained by considering a rigid translation of the surface by a uniform infinitesimal vector $\delta\mathbf{a}$ or a rigid rotation with infinitesimal vector $\delta\boldsymbol{\theta}$. Coordinates on the new surface are obtained by following the normal $\mathbf{n}$ of the original surface. 
 } 
\end{figure}

Using translation invariance of the free energy, we can derive a Gibbs-Duhem relation.
We consider a infinitesimal translation of the surface by a constant vector $\delta \mathbf{a}$. The condition $\partial_i \delta\mathbf{a}=0$ implies using Eq. \ref{AppendixVectorDifferentation}
\begin{eqnarray}
\nabla_i \delta a^j +C_i{}^j  \delta a_n =0,\label{RigidTranslationGibbsDuhem1}\\
\partial_i \delta a_n-C_{ij}\delta a^j =0.\label{RigidTranslationGibbsDuhem2}
\end{eqnarray}
 During translation, we reparametrize the new surface such that each point $(s^1,s^2)$ moves normal to the original surface on the new translated surface (Figure \ref{fig:schematicGibbsDuhem}). Translation invariance then implies the relation (see Appendix \ref{AppendixGibbsDuhem})
\begin{eqnarray}
\label{GibbsDuhemTranslation}
\nabla_j\left[(f_0-\mu^{\alpha} c^{\alpha} )g_{i}{}^{j}-K^{jk} C_{ik}\right]+C_{ik} \nabla_j K^{jk} =-c^{\alpha}\partial_i \mu^{\alpha} .
\end{eqnarray}
Eq. \ref{GibbsDuhemTranslation} is a covariant generalisation for surfaces of the Gibbs-Duhem relation for a three-dimensional multi-component fluid \cite{lomholt2005general,joanny2007hydrodynamic}, with an additional term arising from the passive bending moment tensor.
\subsubsection{Rotation invariance}
We can derive a generalised Gibbs-Duhem relation describing torque balances using infinitesimal rotation described by the pseudo vector $\delta\boldsymbol{ \theta}$, such that the surface is deformed as:
\begin{equation}
\delta X_{\alpha}=\epsilon_{\alpha\beta\gamma}\delta \theta_{\beta} X_{\gamma}.
\end{equation}
The deformation defines a new surface $\mathbf{X}'=\mathbf{X}+\delta\mathbf{X}$, which is reparametrized such that $(s^1,s^2)$ is constant along the normal to the original surface. Rotation invariance then implies (see Appendix \ref{AppendixGibbsDuhem}):
\begin{equation}
\label{GibbsDuhemRotation}
K^{ij}\epsilon_{jk}C_i{}^k=0.
\end{equation}
implying that the tensor $K^{ij}C_i{}^k$ is symmetric. 
\subsubsection{Equilibrium tensions and torques}

The equilibrium tension and bending moment tensors can be obtained by calculating the change of free energy under a surface deformation and using the expression of the virtual work Eq. \ref{VirtualWork2} (Appendix \ref{PassiveMembrane}). The equilibrium tension and bending moments are given by
\begin{eqnarray}
\bar{t}^{ij}_e&=&(f_0-\mu^{\alpha} c^{\alpha} )g^{ij},\label{FluidEquilibriumInPlaneStressTensor}\\
 \bar{m}_e^{ij}&=&K^{ij},\\
m^i_{n,e}&=&0 \quad .\label{FluidEquilibriumNormalTorqueTensor}
  \end{eqnarray}
  with  $\gamma=f_0-\mu^{\alpha} c^{\alpha}$ the bare membrane surface tension.  Using Eqs \ref{DefinitionInPlaneStressTensor} and \ref{DefinitionInPlaneTorqueTensor},  one also obtains the symmetric part of the equilibrium tension tensor $ t^{ij}_{e,s}=(f_0-\mu^{\alpha} c^{\alpha} )g^{ij}-(K_{k}{}^{i} C^{kj}+K_{k}{}^{j} C^{ki})/2$
and the bending moment tensor $m^{ij}_e=K^{ik}\epsilon_{k}{}^{j} $. Using the tangential torque balance equation \ref{TorqueBalanceTangential} then yields the equilibrium tension $t_{e,n}^i=\nabla_j K^{ji}-\epsilon_j{}^i \Gamma^{{\rm ext},j}$. Using Eq. \ref{GibbsDuhemRotation} and \ref{FluidEquilibriumNormalTorqueTensor}, the normal torque balance equation \ref{TorqueBalanceNormal} yields the equilibrium antisymmetric part of the stress, $\epsilon_{ij} t^{ij}=-\Gamma_n^{\rm{ext}}$. 

Combining the Gibbs-Duhem relation \ref{GibbsDuhemTranslation} and the tangential force balance given by Eq. \ref{ForceBalanceTangential}, taking into account the symmetry relation \ref{GibbsDuhemRotation}, leads to the equilibrium condition relating chemical equilibrium gradients to external forces:
\begin{eqnarray}
\label{EquilibriumTangentialForceBalanceEquation}
 c^{\alpha}\partial_j \mu^{\alpha}&=&f^{\rm{ext}}_j-\frac{1}{2}\epsilon^{i}{}_{j} (\partial_i \Gamma_n^{\rm{ext}})-C_{ij} \epsilon_k{}^i \Gamma^{{\rm ext},k}\nonumber \\
 &=&-c^{\alpha}\left[\partial_j U^{\alpha}+C_{ij} (\partial U^{\alpha}/\partial \mathbf{n})\cdot \mathbf{e}^i \right].
 \end{eqnarray}
In the second line, the external force and torque surface densities derive from a potential $U^{\alpha}(s^1,s^2,\mathbf{n})$ acting on component $\alpha$ (Eqs. \ref{ExternalForceDensityEquilibriumFluidMembrane} and \ref{ExternalTorqueDensityEquilibriumFluidMembrane}). Eq. \ref{EquilibriumTangentialForceBalanceEquation}  shows that one can then introduce the effective chemical potential $\mu_{{\rm eff}}^{\alpha}(s^1,s^2)=\mu^{\alpha}(s^1,s^2) +U^{\alpha}(s^1,s^2,\mathbf{n}(s^1,s^2))$, for which $ c^{\alpha}\partial_i \mu_{{\rm eff}}^{\alpha}=0$. 

The remaining normal force balance equation \ref{ForceBalanceNormal} then provides a shape equation for the equilibrium surface shape. 
  
\subsection{Entropy production rate}
We can now calculate the entropy production rate using the variation of the free energy and the Gibbs-Duhem relation derived above.
We consider a region of surface $\mathcal{S}$ enclosed by a fixed contour $\mathcal{C}$, which can deform in 3 dimensions. The rate of change of the free energy $F$ can be written as (see Appendix \ref{AppendixEntropyProductionRate}):
\begin{widetext}
\begin{eqnarray}
\label{RateDissipationFreeEnergy}
\frac{dF}{dt}
&=&\int_\mathcal{S} dS \bigg[-\left(\overline{t}^{ij}-\overline{t}^{ij}_e\right)v_{ij}- (\bar{m}^{ij}-K^{ij})\frac{D C_{ij}}{Dt} -m_n^i \left(\partial_i \omega_n -C_{ij} \omega^j \right)\nonumber\\
&&+ (\partial_i \mu^{\alpha}) j^{\alpha,i}+\mu^{\alpha} r^{\alpha}+\mu_{tot} ^{\alpha}J_n^{\alpha}
+\mathbf{f}^{\rm{ext}}\cdot\mathbf{v} +\boldsymbol{\Gamma}^{\rm{ext}}\cdot \boldsymbol{\omega}\bigg]\nonumber\\
&&\hspace{3cm}+\int_\mathcal{C} dl\nu_i \bigg[-f v^i-\mu^{\alpha} j^{\alpha,i}+ \mathbf{t}^i \cdot \mathbf{v} +\mathbf{m}^{i} \cdot \boldsymbol{\omega}\bigg], 
\end{eqnarray}
where we have introduced the symmetric in-plane shear tensor $v_{ij}$, the rotational of the flow $\boldsymbol{\omega}=\omega^i \mathbf{e}_i+\omega_n \mathbf{n}$, and the corotational derivative of the curvature tensor:
\begin{eqnarray}
\label{DefinitionSymmetricVelocityGradient}
v_{ij}&=&\frac{1}{2}(\nabla_i v_j+\nabla_j v_i)+C_{ij} v_n,\\
\label{DefinitionFlowRotational}
\boldsymbol{\omega}&=&\epsilon^{ij}(\partial_j v_n-C_{jk} v^k ) \mathbf{e}_i+\frac{1}{2} \epsilon^{ij}(\nabla_i v_j ) \mathbf{n},\\
\label{DefinitionCorotationalCurvatureTensor}
\frac{D C_{ij}}{Dt}&=&-\nabla_i(\partial_j v_n)
-v_n C_{ik} C^{k}{}_{j} +v^k \nabla_k C_{ij}   +\omega_n(\epsilon_i{}^k C_{kj} +\epsilon_j{}^k C_{ki})
\end{eqnarray}
\end{widetext}
Note that the in-plane shear tensor $v_{ij}$ is the sum of a contribution from in-plane flows, equal to the symmetric part of the covariant gradient of flow $\nabla_i v_j$, and a contribution arising from normal flows $v^n$, corresponding to in-plane shear induced by the deformation of the surface in three-dimensions. The vorticity $\boldsymbol{\omega}$ of the flow has a normal part arising from the two-dimensional vorticity of the flow $\epsilon^{ij} \nabla_i v_j/2$, and a tangential part specific to curved surfaces. The bending rate tensor $D C_{ij}/Dt$ has the form of a corotational derivative, with the third term in Eq. \ref{DefinitionCorotationalCurvatureTensor} corresponding to advection of the curvature, and the last two terms to a corotational term. In Eq. \ref{RateDissipationFreeEnergy}, we have not included contributions from the antisymmetric part of $\bar{m}_{ij}$. Note that the bending moment tensor can always be chosen to be symmetric in the force balance equations, see Appendix \ref{AppendixEntropyProductionRate}. 

We can read off the entropy production rate in the surface per unit area from Eq.  \ref{RateDissipationFreeEnergy}:
\begin{eqnarray}
\label{EntropyProduction}
T \pi=\overline{t}^{ij}_d v_{ij}+ \bar{m}_d^{ij}\frac{D C_{ij}}{Dt} +m_n^i \left(  \partial_i \omega_n -C_{ij} \omega^j \right)-(\partial_i \mu^{\alpha}) j^{\alpha,i} - \mu^{\alpha} r^{\alpha},
\end{eqnarray}
where $\overline{t}^{ij}_d=\bar{t}^{ij}-\gamma g^{ij}$ and $\bar{m}^{ij}_d=\bar{m}^{ij}-K^{ij}$ are the dissipative part of the in-plane stress and bending moment tensor. The mechanical contribution to dissipation can be also understood starting from Eq. \ref{VirtualWork2} using $T\pi_m=\delta W_d/\delta t$, where $\delta W_d$ is the work done by dissipative forces, together with Eqs. \ref{LagrangianDerivativeMetric} and \ref{LagrangianDerivativeChristoffel}.  Note that the entropy production rate is a sum of products of conjugate thermodynamic fluxes and forces,  which all vanish at thermodynamic equilibrium. The pairs of conjugate fluxes and forces are listed in Table \ref{FluxesForcesTable}.

We now briefly discuss the conjugate fluxes and forces. The dissipative in-plane tension tensor $\overline{t}^{ij}_d$ is conjugate to the in-plane shear rate $v_{ij}$, corresponding to the dissipative cost of introducing in-plane deformations in the surface. The coupling between the in-plane dissipative bending moment $\overline{m}_d^{ij}$ and the bending rate tensor $DC_{ij}/Dt$ arises only for curved surfaces and is associated to the dissipative cost of changing the surface shape in three dimensions. The coupling between the normal moment $m_n^i$ and the vorticity gradient of flow $(\partial_i \boldsymbol{\omega})\cdot\mathbf{n}=\partial_i \omega_n-C_{ij}\omega^j$ is a generalisation to curved surface of a coupling which also arises for planar surfaces, and is associated to the dissipative cost of gradients of rotations within the surface \cite{furthauer2012active}. Finally, the two last terms in Eq. \ref{EntropyProduction} correspond to couplings of the chemical potential and its gradient to the rates of reactions and the flux of diffusion of species in the surface \cite{joanny2007hydrodynamic}.
 
The flux of free energy entering the surface from the adjacent bulk reads
\begin{equation}
J_n^f=\mathbf{f}^{\rm{ext}}\cdot\mathbf{v}+\boldsymbol{\Gamma}^{\rm{ext}}\cdot \boldsymbol{\omega}+\mu_{tot}^{\alpha} J_n^{\alpha},
\end{equation}
which corresponds to the sum of the mechanical power acting on the surface and of the influx of chemical energy in the surface.
The flux of free energy tangential to the surface reads:
\begin{equation}
J^{f,i}=f v^i -\mathbf{t}^i \cdot \mathbf{v}  -\mathbf{m}^i \cdot \boldsymbol{\omega}+\mu^{\alpha} j^{\alpha, i}
\end{equation}
where $f v^i$ is the advection of free energy, $\mu^{\alpha} j^{\alpha,i}$ is the flux of chemical free energy, and the remaining terms describe the mechanical power tangential to the surface at its boundaries. 

\begin{widetext}
\begin{center}
\begin{table}
\centering
  \begin{tabular}{|c|c|}\hline
  Flux & Force \\\hline
  In-plane shear tensor $v_{ij}$ & In-plane tension tensor ${\bar{t}}^{ij}_d$\\\hline
Bending rate tensor $\frac{ D C_{ij}}{Dt}$& In-plane bending moment tensor ${\bar{m}}_d^{ij}$ \\\hline
Vorticity gradient $(\partial_i \boldsymbol{\omega})\cdot\mathbf{n}$ & Normal moment $m_n^i$ \\\hline
Diffusion flux  $j^{i,\alpha}$ &Chemical potential gradient $-\partial_i \mu^{\alpha}$\\\hline
Chemical reaction rate $r^{\alpha}$ & Chemical potential $\mu^{\alpha}$\\\hline
  \end{tabular}
  \caption{List of pairs of conjugate thermodynamics fluxes and forces in a thin active surface. }\label{FluxesForcesTable}
\end{table}
\end{center}
\end{widetext}

\begin{figure}[h!]
 \centering
  \includegraphics[width=8cm]{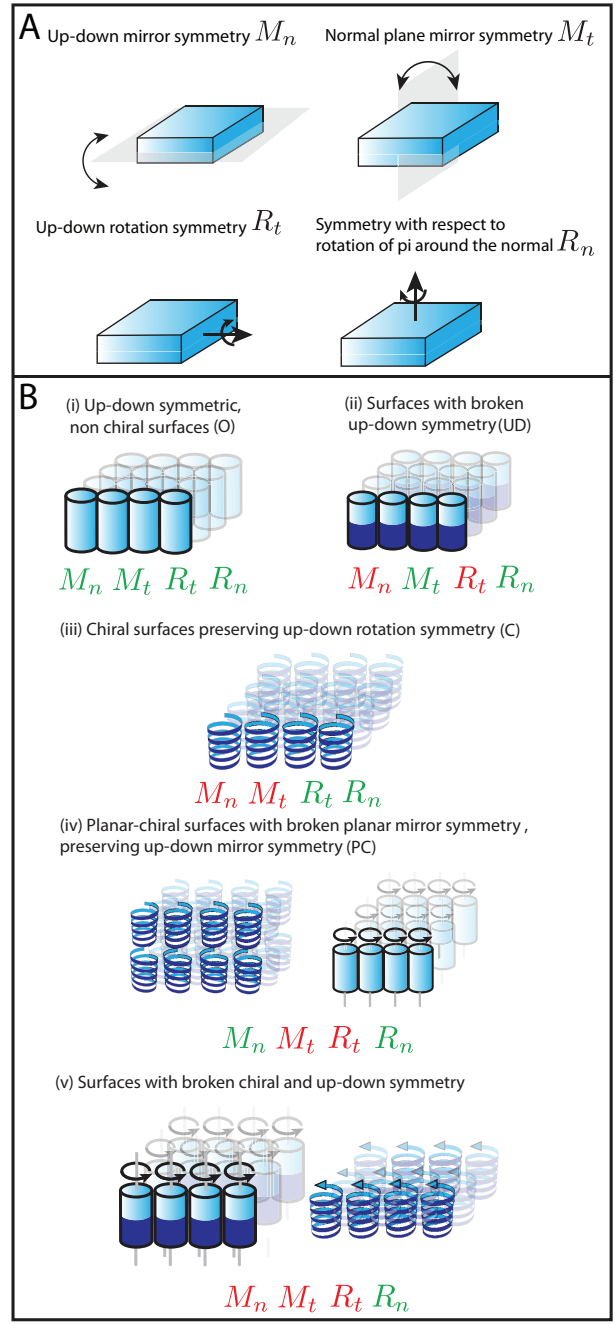}
 \caption{
 \label{fig:symmetries}Classification of surfaces with in-plane rotation symmetry. A. The surface state can change under up-down mirror symmetry $M_n$,  mirror symmetry $M_t$, up-down rotation symmetry $R_t$, and rotation by $\pi$ around the normal $R_n$. The symmetry $R_n$ is not broken for a surface with in-plane rotation symmetry. B. Surfaces with in-plane rotation symmetry can be categorised in 5 classes according to their symmetries. Schematics give examples of actual surfaces belonging to each category. Red and green letters indicate respectively broken and preserved symmetries.} 
\end{figure}

\subsection{Mirror and rotation symmetries of surfaces}

Constitutive relations describing the active surface must respect the symmetries satisfied by the surface \cite{curie1894symmetry}. We therefore classify surfaces by asking whether the state of an element of surface is preserved under application of symmetries (Fig. \ref{fig:symmetries}). 

We restrict ourselves to surfaces with rotation symmetry in the plane. We then find that that 3 sets of discrete symmetries can be associated to thin shells: up-down mirror symmetry $M_n$, mirror symmetry with respect to a plane normal to the surface $M_t$,  and up-down rotation symmetry $R_t$ (Fig. \ref{fig:symmetries}A).  $M_t$ corresponds to a mirror symmetry by a normal plane going along an arbitrary tangent vector $\mathbf{t}$, $R_t$ to a rotation of $\pi$ around an arbitrary tangent vector $\mathbf{t}$.  The corresponding transformations rules are given in Appendix \ref{AppendixSurfaceSymmetries}. Because inversion of space can be written as the combination of $M_n$ and the rotation of $\pi$ around the normal $R_n$, inversion of space and $M_n$ are broken or preserved simultaneously for a surface with in-plane rotation symmetry. Furthermore, combination of two of the symmetries $M_n$, $M_t$ and $R_t$ yield the third one, such that at least two of these symmetries must be broken. As a result, surfaces can be classified into 5 different classes:  (i) up-down symmetric, non-chiral surfaces (type 0) preserve all three symmetries, (ii) non-chiral surfaces with broken up-down symmetry (type UD) preserve $M_t$ but break $M_n$ and $R_t$, (iii) chiral surfaces with up-down rotation symmetry break all mirror symmetries $M_t$ and $M_n$ but preserve $R_t$ (type C) (iv) planar-chiral surfaces preserve up-down mirror symmetry $M_n$ but break $M_t$ and $R_t$ (type PC) (v) up-down asymmetric and chiral surfaces break $M_n$, $M_t$ and $R_t$ (Fig. \ref{fig:symmetries}). Note that we choose to denote surfaces breaking $M_t$ and not $M_n$ planar-chiral surfaces because they break mirror-symmetry in the plane, but these surfaces are not necessarily made of chiral molecules (Fig. \ref{fig:symmetries}B).

\subsection{Constitutive and hydrodynamic equations}

Using the conjugate thermodynamic forces and fluxes obtained from Eq. \ref{EntropyProduction} and listed in Table \ref{FluxesForcesTable}, we write a generic linear response theory taking into account the symmetries of an active fluid surface. For simplicity, we consider that a single chemical reaction occurs in the surface converting a fuel species $F$ into a product species $P$. The fuel and product species have the same mass. We denote $\Delta\mu=\mu^F-\mu^P$ the difference of chemical potential between the field and product species, $r=-r^F=r^P$ the rate of fuel consumption and $\mathbf{j}=\mathbf{j}^F=-\mathbf{j}^P$ its flux. We also assume that no chemical exchange exists between the membrane and its surrounding, such that the normal fluxes $J_n^{\alpha}$ and $J_n^{\rho}$ vanish. In the linear response theory, we expand the tensors $\bar{t}^{ij}_d$, $\bar{m}^{ij}_d$, $m_n^i$, diffusion flux $j^{i,\alpha}$ and chemical reaction rate $r^{\alpha}$ to linear order in the rates of deformation $v_{ij}$, $DC_{i}{}^{j}/Dt$, $(\partial_i\boldsymbol{\omega})\cdot\mathbf{n}$, chemical potential $\Delta\mu$ and chemical potential gradient $\partial_i \Delta\mu$.

The stress and moment tensor can then be decomposed as
\begin{eqnarray}
\overline{t}^{ij}_d&=&\overline{t}^{ij}_0+\overline{t}^{ij}_{\rm{UD}}+\overline{t}^{ij}_{\rm{C}}+\overline{t}^{ij}_{\rm{PC}},\nonumber\\
\overline{m}^{ij}_d&=&\overline{m}^{ij}_0+\overline{m}^{ij}_{\rm{UD}}+\overline{m}^{ij}_{\rm{C}}+\overline{m}^{ij}_{\rm{PC}},\nonumber\\
m^i_n&=&m^i_{n0}+m^i_{n\rm{UD}}+m^i_{n\rm{C}}+m^i_{n\rm{PC}},\label{GeneralTensorDecompositionSymmetry}
\end{eqnarray}
where $\overline{t}^{ij}_0$ is the part of the stress tensor that exists for any surface, $\overline{t}^{ij}_{\rm{UD}}$ correspond to terms present when the surface breaks up-down symmetry, $\overline{t}^{ij}_{\rm{C}}$ exist for chiral surfaces, and $\overline{t}^{ij}_{\rm{PC}}$ for planar-chiral surfaces. Similar rules apply for the decomposition of the bending moment tensor and normal moment tensor.

To express constitutive equations for each of the components, we then write all possible terms of the expansion of the generalised forces in the fluxes at linear order, and ask whether the corresponding terms break the symmetry $M_n$, $M_t$, $R_t$ according to the signatures given in Appendix \ref{AppendixSurfaceSymmetries}. The contributions to the stress tensor then read
\begin{widetext}
\begin{eqnarray}
\overline{t}^{ij}_0&=&2\eta \tilde{v}^{ij} +\eta_b v_k{}^k g^{ij}+\zeta g^{ij} \Delta \mu\nonumber\\
\overline{t}^{ij}_{\rm{UD}}&=&2 \bar{\eta} \frac{D \tilde{C}^{ij}}{Dt}+\bar{\eta}_b \frac{D C_k{}^k}{Dt} g^{ij}+2 \tilde{\zeta}\tilde{C}^{ij} \Delta \mu+\zeta' C_k{}^k g^{ij} \Delta\mu \nonumber \\
\overline{t}^{ij}_{\rm{C}}&=&\eta_{\rm{C}} \left(\epsilon^{ik} \frac{D C_{k}{}^{j}}{Dt}+\epsilon^{jk} \frac{DC_{k}{}^{i}}{Dt}\right)+\zeta_{\rm{C}} \left(\epsilon^i{}_k C^{kj}+\epsilon^j{}_k C^{ki}\right)\Delta\mu\nonumber\\
\overline{t}^{ij}_{\rm{PC}}&=&\eta_{\rm{PC}} \left(\epsilon^i{}_k v^{kj}+\epsilon^j{}_k v^{ki}\right),
\label{ConstitutiveEquationtij}
\end{eqnarray}
where we have introduced the notation $\tilde{A}_{ij}=A_{ij}-\frac{1}{2}A_{k}{}^k g_{ij}$ for the traceless part of a tensor $\mathbf{A}$.
The moment tensor reads
\begin{eqnarray}
\overline{m}^{ij}_0&=&2 \eta_c \frac{D \tilde{C}^{ij}}{Dt}+  \eta_{cb} \frac{D C_{k}{}^{k}}{Dt}g^{ij}+2 \tilde{\zeta_c} \tilde{C}^{ij} \Delta \mu+\zeta_c' C_k{}^k g^{ij} \Delta\mu   \nonumber \\
\overline{m}^{ij}_{\rm{UD}}&=&2\bar{\eta} \tilde{v}^{ij}+\bar{\eta}_b v_k{}^k g^{ij}+ \zeta_c g^{ij} \Delta \mu \nonumber \\
\overline{m}^{ij}_{\rm{C}}&=&-\eta_{\rm{C}} \left(\epsilon^i{}_k v^{kj}+\epsilon^j{}_k v^{ki}\right)\nonumber\\
\overline{m}^{ij}_{\rm{PC}}&=&\eta_{\rm{cPC}} \left(\epsilon^{ik} \frac{D C_{k}{}^{j}}{Dt}+\epsilon^{jk} \frac{DC_{k}{}^{i}}{Dt}\right)+\zeta_{\rm{PC}} \left(\epsilon^i{}_k C^{kj}+\epsilon^j{}_k C^{ki}\right)\Delta \mu.
 \label{ConstitutiveEquationmij}
\end{eqnarray}
In Eq. \ref{ConstitutiveEquationmij}, we have only introduced symmetric contributions to the bending moment tensor.  The normal moment reads
\begin{eqnarray}
m^i_{n0}&=&\lambda (\partial^i \omega_n -C^{ij} \omega_j)+ \chi\epsilon^{ij} \partial_j \Delta\mu \nonumber\\
m^i_{n\rm{UD}}&=&0\nonumber\\
m^i_{n\rm{C}}&=& \chi_{\rm{C}} C^{ij} \partial_j \Delta\mu\nonumber\\
m^i_{n\rm{PC}}&=&\lambda_{\rm{PC}} \epsilon^{ij} (\partial_j \omega_n -C_{jk} \omega^k)+ \chi_{\rm{PC}}  \partial^i \Delta\mu.
\label{ConstitutiveEquationmn}
\end{eqnarray}
The rate of fuel consumption then reads
\begin{eqnarray}
r&=&-(\zeta+\zeta' C_k{}^k) v_{k}{}^k -2\tilde{\zeta} \tilde{C}^{ij}\tilde{v}_{ij}-2\zeta_{\rm{C}} \epsilon^i{}_k C^{kj} v_{ij}\nonumber\\
&&-(\zeta_c+\zeta_c' C_k{}^k) \frac{D C_{k}{}^k}{Dt}-2 \tilde{\zeta_c}\tilde{C}^{ij}\frac{D \tilde{C}_{ij}}{Dt}-2\zeta_{\rm{PC}} \epsilon^i{}_k C^{kj} \frac{DC_{ij}}{Dt}+\Lambda \Delta \mu \label{ConstitutiveEquationr}.
\end{eqnarray}
and the fuel flux relative to the centre of mass is given by
\begin{equation}
j^{i}=-L\partial^i \Delta\mu + \left(\chi  \epsilon^{ij}+\chi_{\rm{C}} C^{ij}+\chi_{\rm{PC}} g^{ij}\right) (\partial_j \omega_n -C_{jk} \omega^k) \label{ConstitutiveEquationj}.
\end{equation}
\end{widetext}

$\eta$, $\eta_b$, $\bar{\eta}$, $\bar{\eta}_b$, $\eta_c$, $\eta_{cb}$, $\eta_{\rm{C}}$, $\lambda$, $\Lambda$  and $L$ are dissipative couplings, $\zeta$, $\zeta'$, $\tilde{\zeta}$, $\zeta_c$, $\tilde{\zeta}_c$, $\zeta'_c$, $\chi$, $\chi_{\rm{C}}$ and $\chi_{\rm{PC}}$ are reactive couplings. The viscosities depend in general on the curvature tensor $C_{ij}$; here we have not taken this dependency into account for simplicity. We have introduced terms proportionals to $\eta_{\rm{PC}}$, $\eta_{\rm{cPC}}$ and $\lambda_{\rm{PC}}$ corresponding to odd or Hall viscosities which do not contribute to dissipation. These are reactive coefficients, and the time signatures of the constitutive equations imply that they change sign under time reversal, which could exist for example in the presence of a magnetic field \cite{avron1998odd}.  Active tensions and bending moments proportional to the difference of chemical potential $\Delta \mu$ depend on the curvature tensor. In the constitutive equations \ref{ConstitutiveEquationtij}-\ref{ConstitutiveEquationj}, we have expanded these terms to first order in the curvature tensor $C_{ij}$. Although we have not written explicitly this dependency here, the phenomenological coefficients also depend in general in the concentration fields $c^{\alpha}$. Positivity of entropy productions implies that the viscosities $\eta$, $\eta_b$, $\eta_c$, $\eta_{cb}$, and $\lambda$ are positive, however the up-down asymmetric viscosities $\bar{\eta}$, $\bar{\eta}_b$ and chiral viscosity $\eta_{\rm{C}}$ can be positive or negative.  

In the equations above, the contribution to the two-dimensional stress $\bar{t}_0^{ij}$ is the generalisation for curved surfaces of the generic hydrodynamic equations of a three-dimensional active gel \cite{kruse2005generic}: $\eta$ and $\eta_b$ are respectively the planar shear and bulk viscosity of the surface, and $\zeta\Delta\mu$ is an active tension arising in the surface from active processes. Additional viscous tensions proportional to $\bar{\eta}$, $\bar{\eta}_b$, and $\eta_C$ arise for a curved surface due to the dissipative cost of changing the surface curvature. We also find new active terms for the tension tensor of a curved surface proportional to $\tilde{\zeta}$, $\zeta'$, $\zeta_C$, that depend on the curvature tensor $C_{ij}$. In particular, anisotropic active stresses can arise in a curved surface isotropic in the plane, due to the anisotropy of the curvature.

 Active terms for the moment tensor introduced in Eq. \ref{ConstitutiveEquationmij} are specific to thin films and correspond to actively induced torques in the film. The active torque $\zeta_c$, arising in a surface with broken up-down symmetry, can induce active bending of a flat surface.

Combining the constitutive equations \ref{ConstitutiveEquationtij}-\ref{ConstitutiveEquationj}, the force and torque balance equations \ref{ForceBalanceEquation} and \ref{TorqueBalanceEquation}, and the concentration balance equations \ref{ConcentrationBalanceEquation} yield dynamic equations for the surface shape, the velocity field on the surface $\mathbf{v}$ and the concentration fields on the surface $c^k$. While the constitutive equations obtained here are linear, the dynamics equations for the surface shape are non-linear due to geometric couplings.

\subsection{Instabilities of a homogeneous active Helfrich membrane}

In this section, we restrict ourselves to non-chiral surfaces with broken up-down symmetry and discuss low Reynolds numbers where inertial terms can be neglected. Starting from a description of a passive surface with the Helfrich free energy, we consider effects introduced by additional active terms.

\subsubsection{General equations}
A passive fluid membrane described by the Helfrich energy with membrane tension $\gamma$, bending modulus $\kappa$, gaussian bending modulus $\kappa_g$ and spontaneous curvature $C_0$ has the equilibrium tension and bending moment tensor (Appendix \ref{PassiveMembrane})
\begin{eqnarray}
\bar{t}_{ij}&=&\big(\gamma+(\kappa+\kappa_g) (C_k{}^k)^2-4\kappa C_k{}^k C_0-\kappa_g C_{l}{}^{k}C_{k}{}^{l}\big)g_{ij} ,\\
\bar{m}_{i}{}^{j}&=&\left(2(\kappa+\kappa_g) C_k{}^k -4\kappa C_0\right) g_i{}^j- 2\kappa_g C_i{}^j.
\end{eqnarray}
Starting from such a passive fluid membrane, the constitutive relation for the tension and bending moment tensor of an active surface reads, neglecting viscous terms for this discussion and only keeping terms to first order in the curvature:
\begin{widetext}
\begin{eqnarray}
\overline{t}^{ij}&\simeq&\left(\gamma+\zeta\Delta\mu+(-4\kappa C_0 +\zeta' \Delta\mu)  C_k{}^k\right) g^{ij} +2 \tilde{\zeta} \Delta \mu   \tilde{C}^{ij} \label{ConstitutiveEquationtij2},\\
\bar{m}^{ij}&\simeq&\left(\left(2\kappa+2\kappa_g+\left(\zeta_c'-\tilde{\zeta}_c \right)\Delta\mu\right) C_k{}^k+\left(\zeta_c \Delta\mu-4\kappa C_0\right) \right) g^{ij}  -2\left(\kappa_g- \tilde{\zeta_c}\Delta \mu \right) C^{ij}. \label{ConstitutiveEquationmij2}
\end{eqnarray}
\end{widetext}
Introducing a surface tension renormalised by activity $\bar{\gamma}=\gamma+\zeta\Delta\mu$, and similarly the renormalized bending moduli $\bar{\kappa}_g=\kappa_g-\tilde{\zeta_c}\Delta\mu$, $\bar{\kappa}=\kappa+(\tilde{\zeta_c}+\zeta'_c)\Delta\mu/2$ and spontaneous curvature $\bar{C}_0=\kappa C_0/\bar{\kappa}-\zeta_c\Delta\mu/4\bar{\kappa}$, one obtains
\begin{widetext}
\begin{eqnarray}
\overline{t}^{ij}&=&\left(\bar{\gamma}+(-4\bar{\kappa} \bar{C}_0 +(\zeta'-\zeta_c) \Delta\mu)  C_k{}^k\right) g^{ij} + 2\tilde{\zeta} \Delta \mu \tilde{C}^{ij} \label{ConstitutiveEquationtij3},\\
\bar{m}^{ij}&=&\left(\left(2\bar{\kappa}+2\bar{\kappa}_g\right) C_k{}^k -4\bar{\kappa} \bar{C}_0\right) g^{ij}  -2\bar{\kappa}_g  C^{ij}. \label{ConstitutiveEquationmij3}
\end{eqnarray}
\end{widetext}

Two active terms proportional to $\Delta\mu$ remain in the constitutive equation \ref{ConstitutiveEquationtij3}. Active terms therefore do not simply renormalise the physical parameters of the Helfrich membrane, but introduce other physical effects. To clarify the role of these terms, we discuss below simple surface geometries of active Helfrich membranes and show that they can result in instabilities of a flat surface.

\subsubsection{Instabilities of a flat surface}
\label{FlatSurfaceInstability}
\begin{figure}[ht!!]
 \centering
  \includegraphics[width=9cm]{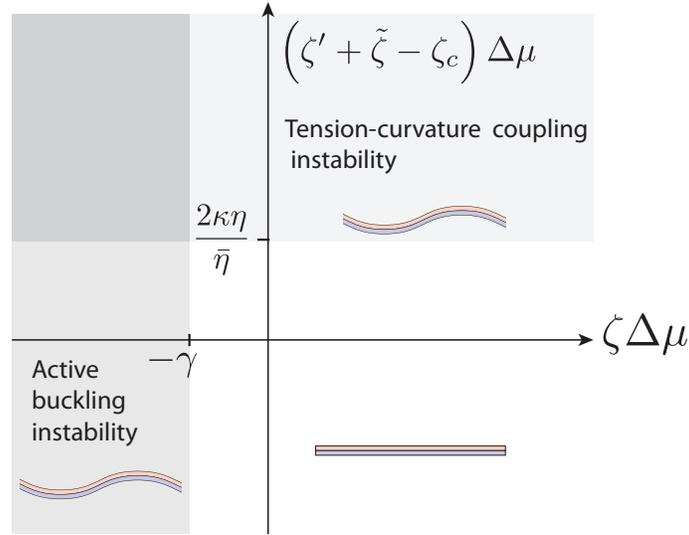}
 \caption{
 \label{fig:instability}Phase diagram for the stability of an flat active Helfrich membrane with up-down asymmetry, as a function of the active tension $\zeta\Delta\mu$ and the active tension-curvature coupling term $(\zeta'+\tilde{\zeta}-\zeta_c) \Delta \mu$. For simplicity, we consider here the case $\tilde{\zeta}_c=\zeta_c'=0$.} 
\end{figure}

We consider here a flat, homogeneous and compressible membrane. We ignore here the surrounding medium and the membrane is therefore free from external forces and torques. Perturbations of the flat shape are described in the Monge gauge by the height $h(x,y)$, such that the surface position is given by $\mathbf{X}(x,y)=x\mathbf{u}_x+y\mathbf{u}_y+h(x,y)\mathbf{u}_z$. We take here for simplicity the bulk viscosities $\eta_b=\eta_{bc}=0$ and we obtain the shape equation (Appendix \ref{LinearStability})

\begin{eqnarray}
 q^4\bigg[(\eta_c-\frac{\bar{\eta}^2}{\eta}) \partial_t +2\kappa +(\tilde{\zeta_c}+\zeta'_c)\Delta\mu+\frac{\bar{\eta}(\zeta_c-\zeta' -\tilde{\zeta})\Delta \mu}{\eta}\bigg]\tilde{h} +q^2(\gamma+\zeta\Delta\mu) \tilde{h}=0,
\end{eqnarray}
where we have introduced the Fourier transform of the height, $\tilde{h}(q_x,q_y)=\frac{1}{2\pi}\int dx dy h(x,y) e^{-i (q_x x+ q_y y)}$, and $q=\sqrt{q_x^2+q_y^2}$.
The second law of thermodynamics imposes that $\eta_c \eta>\bar{\eta}^2$. We find that the active flat surface undergoes shape instabilities for (Figure \ref{fig:instability}):
\begin{eqnarray}
\zeta\Delta\mu&<&-\gamma,\label{FluidInstability1}\\
\frac{\bar{\eta}(\zeta' +\tilde{\zeta}-\zeta_c)\Delta \mu}{\eta}&>&2\kappa+(\tilde{\zeta}_c+\zeta'_c)\Delta\mu \label{FluidInstability2}.
\end{eqnarray}
The first condition corresponds to the classical buckling instability occurring when active stresses are compressive and establish a negative surface tension $\bar{\gamma}=\gamma+\zeta\Delta\mu<0$ in the membrane. 

In the second condition, the instability is favoured by negative values of $\tilde{\zeta}_c+\zeta'_c$ or positive values of $\bar{\eta}(\zeta'+\tilde{\zeta}-\zeta_c)$. Negative values of $\tilde{\zeta}_c+\zeta'_c$ lower the effective bending modulus $\bar{\kappa}$. Positive values of $\bar{\eta}(\zeta'+\tilde{\zeta}-\zeta_c)$ induce an instability coupling the membrane shape to tangential flows. This instability can be understood from the dependency of the tension on curvature (Eqs. \ref{DefinitionInPlaneStressTensor} and \ref{ConstitutiveEquationtij}). Because of this dependency, a perturbation of the surface shape results in regions of low and high surface tension, depending on the sign of the local mean curvature and the sign of the coefficient $\zeta'+\tilde{\zeta}-\zeta_c$ which couples the tension tensor to the curvature tensor. Differences of surface tensions result in flows towards region of higher surface tension. These flows generate further in-plane torques when the surface has a non-zero up-down asymmetric viscosity $\bar{\eta}$ or $\bar{\eta}_b$. A shape instability occurs when the sign of this additional torque leads to further deformation of the surface.
\section{Active elastic thin shell}

In addition to fluid surfaces, the formalism presented here can also be used for elastic surfaces. We discuss here isotropic active elastic thin shells.
\subsection{Hookean elasticity}
We first write generic constitutive equations for a Hookean elastic shell. Rather than inferring tensions and moment tensors from three-dimensional stresses, we directly obtain generic two-dimensional constitutive equations \cite{koiter1973foundations, koiter1970mathematical}. We consider a surface with reference shape $\mathbf{X}$ and a deformation field $\mathbf{u}$, such that the deformed surface has position $\mathbf{X'}=\mathbf{X}+\mathbf{u}$. Using the differential virtual work expression Eq. \ref{VirtualWork2}, the change of virtual work induced by the deformation field $\mathbf{u}$ reads to first order in the deformation field:
\begin{equation}
\label{Work}
\Delta W\simeq\int_{\mathcal{S}} dS \left[\bar{t}^{ij} u_{ij} + \bar{m}^{ij} c_{ij}+m_n^i\Omega_i\right]
\end{equation}
where we have introduce the deformation tensors $u_{ij}=\Delta g_{ij}/2$, $c_{ij}= (g_{ik} \Delta C_{j}{}^{k}+g_{jk} \Delta C_{i}{}^{k})/2 $ and $\Omega_i=\epsilon^{j}{}_{k} \Delta \Gamma_{ij}^k/2$, and we have assumed as for the fluid case that $\bar{m}_{ij}$ can be taken to be symmetric. The deformation tensors read to first order in the deformation field:
\begin{eqnarray}
u_{ij}&=&\frac{1}{2}\left(\nabla_i u_j+\nabla_j u_i\right) + C_{ij} u^n,\\
c_{ij}&=&-\nabla_i(\partial_j u_n)-u_n C_{ik} C^{k}{}_{j} +(\nabla_k C_{ij}) u^k+\frac{1}{2} \epsilon_{kl} \nabla^k u^l( \epsilon_{i}{}^k C_{ki} +\epsilon_{j}{}^{k} C_{ki}) ,\\
\Omega_i&=&\frac{1}{2}\nabla_i(\epsilon^j{}_k \nabla_j u^k)- C_{ij}\epsilon^{jk} (\partial_k u_n -C_{kl} u^l).
\end{eqnarray}
where we have used Eqs. \ref{VariationMetricCovariant}, \ref{VariationCurvatureTensor} and \ref{GradientRotationTensor}. We can then identify that the in-plane stress $\bar{t}_{ij}$, in-plane bending moments $\bar{m}_{ij}$, and normal moments $m_n$ are conjugate to the deformation tensor $u_{ij}$, $c_{ij}$ and $\Omega_i$. We can therefore express Hookean elasticity by the following constitutive relations:
\begin{eqnarray}
\bar{t}_{ij}^e&=&E_{ijkl} u^{kl} +G_{ijkl} c^{kl}, \\
{\bar{m}}_{ij}^e&=&K_{ijkl} u^{kl} +F_{ijkl} c^{kl},\\
m_n^i&=&H^{ij} \Omega_j,\label{ElasticConstitutiveEquation}
\end{eqnarray}
where we have introduced the Hookean elastic moduli tensors $E$, $F$, $G$, $K$ and $H$. For a shell in thermodynamic equilibrium with free energy $F$, $G_{ijkl}=\partial^2 F/\partial c_{kl}\partial u_{ij}=\partial^2 F/\partial u_{ij}\partial c_{kl}=K_{klij}$ and $H_{ij}=\partial^2 F/\partial \Omega_i\partial \Omega_j=\partial^2 F/\partial \Omega_j\partial\Omega_i= H_{ji}$. On an homogeneous elastic shell, the metric, curvature and Levi-Civita tensors can be used to define these elasticity tensors. We therefore simplify the general relations above in the form
\begin{eqnarray}
\overline{t}^{ij}_0&=&E_1 u^{ij} +E_2 u_k{}^k g^{ij}\nonumber\\
\overline{t}^{ij}_{\rm{UD}}&=&G_1 c^{ij}+G_2 c_k{}^k g^{ij} \nonumber \\
\overline{t}^{ij}_{\rm{C}}&=&G_{\rm{C}} \left[\epsilon^i{}_k c^{kj}+\epsilon^j{}_k c^{ki}\right]\nonumber\\
\overline{t}^{ij}_{\rm{PC}}&=&E_{\rm{PC}} \left[\epsilon^i{}_k u^{kj}+\epsilon^j{}_k u^{ki}\right],
\label{ElasticConstitutiveEquationtij}
\end{eqnarray}
\begin{eqnarray}
\overline{m}^{ij}_0&=&F_1 c^{ij}+  F_2 c_{k}{}^{k}g^{ij} \nonumber \\
\overline{m}^{ij}_{\rm{UD}}&=&K_1 u^{ij}+K_2 u_k{}^k g^{ij} \nonumber \\
\overline{m}^{ij}_{\rm{C}}&=&-K_{\rm{C}}  \left[\epsilon^i{}_k u^{kj}+\epsilon^j{}_k u^{ki}\right]\nonumber\\
\overline{m}^{ij}_{\rm{PC}}&=&F_{\rm{PC}} \left[\epsilon^i{}_k c^{kj}+\epsilon^j{}_k c^{ki}\right],
\label{ElasticConstitutiveEquationmij}
\end{eqnarray}
\begin{eqnarray}
m^i_{n0}&=&H_1 \Omega^i\nonumber\\
m^i_{n\rm{UD}}&=&0\nonumber\\
m^i_{n\rm{C}}&=& H_{\rm{C}} \epsilon^{ik}C_{kj} \Omega^j+\bar{H}_{\rm{C}} \epsilon_{jk}C^{ki} \Omega^j \nonumber\\
m^i_{n\rm{PC}}&=&H_{\rm{PC}}\epsilon^{ij} \Omega_j.
\label{ElasticConstitutiveEquationmn}
\end{eqnarray}
where we have decomposed the tension and bending moment tensors according to the symmetry class of the shell, as in the fluid case (Eqs. \ref{GeneralTensorDecompositionSymmetry}).
The coefficient $E_k$, $F_k$, $G_k$, $K_k$ and $H_k$ are elastic moduli, and we have written all terms allowed by symmetry for an homogeneous isotropic elastic material, at lowest order in the curvature tensor $C_{ij}$. For an elastic shell at thermodynamic equilibrium, $G_1=K_1$, $G_2=K_2$, $G_C=K_C$, $H_C=\bar{H}_C$, as a result of the tensor symmetries $G_{ijkl}=K_{klij}$ and $H_{ij}=H_{ji}$ (see after Eq. \ref{ElasticConstitutiveEquation}). A linear shell theory for a homogeneous elastic shell yields $E_1=Eh/(1+\nu)$, $E_2=Eh\nu/(1-\nu^2)$, $F_1=E h^3/(12(1+\nu))$, $F_2=E h^3\nu/(12(1-\nu^2))$ and other coefficients equal to zero, with $E,\nu $ the 3D elastic modulus and Poisson ratio of the shell material and $h$ the thickness of the shell \cite{koiter1970mathematical, reddy2006theory}. The elastic moduli $E_{\rm{PC}}$, $F_{\rm{PC}}$ and $H_{\rm{PC}}$ do not contribute to the work Eq. \ref{VirtualWork2} and only exist for non-equilibrium systems: they vanish for an elastic shell at equilibrium as they do not derive from a free energy.
\subsection{Constitutive relations for an active elastic shell}

For an active elastic shell, an active contribution to the tension and bending moment tensors can be added to the elastic contribution in Eqs. \ref{ElasticConstitutiveEquationtij}-\ref{ElasticConstitutiveEquationmn}:
\begin{eqnarray}
\bar{t}_{ij}^a&=&\left[\zeta g^{ij} +\zeta' C_k{}^k g^{ij}+2 \tilde{\zeta} \tilde{C}^{ij}\right] \Delta \mu,\\
\bar{m}_{ij}^a&=&\left[\zeta_c g^{ij} +\zeta_c' C_k{}^k g^{ij}+2 \tilde{\zeta_c}\tilde{C}^{ij} \right] \Delta \mu ,
\end{eqnarray}
where we restrict ourselves here for simplicity to the case of an non-chiral surface with broken up-down symmetry. In the expansions above, the terms proportional to $\zeta$ and $\zeta_c$, which are to lowest order in the curvature, introduce respectively an active tension and an active torque within the elastic shell.  

We can perform a stability analysis of an elastic flat active surface, similar to the calculation of section \ref{FlatSurfaceInstability} for the fluid case (Appendix \ref{LinearStability}). We find 
\begin{align}
\xi \partial_t \tilde{h} &=-\zeta\Delta\mu q^2 \tilde{h}-\left[F+\zeta_c'+\tilde{\zeta}_c -\frac{K}{E}\left(G+(\zeta'+\tilde{\zeta}-\zeta_c)\Delta\mu\right)\right]q^4  \tilde{h} .
\end{align}
where we have introduced an effective external friction force normal to the surface, with friction coefficient $\xi$, the Fourier transform of the height $\tilde{h}$, and the coefficients $F=F_1+F_2$, $K=K_1+K_2$, $E=E_1+E_2$ and $G=G_1+G_2$. The elastic surface is then unstable for
\begin{align}
\zeta\Delta\mu&<0\\
\frac{K(\zeta'+\tilde{\zeta}-\zeta_c)\Delta\mu}{E}&>F-\frac{KG}{E}+(\zeta_c'+\tilde{\zeta}_c)\Delta\mu.
\end{align}
As for the fluid case (Eqs. \ref{FluidInstability1}-\ref{FluidInstability2}), an instability can arise from active compressive stresses in the surface, or from active couplings between tension and curvature.

The deformation induced by a gradient of active stress and torques in a cylindrical elastic shell have been discussed in Ref. \cite{berthoumieux2014active}. In this work, it was shown that deformation profiles depend on two characteristic lengths which depend on the shell bending modulus, elastic modulus, cylinder radius and on the active tension acting within the shell.

\section{Discussion}

We have developed a general, covariant theory for the dynamics of active surfaces. Starting from balances of forces and torques, we have derived an expression for the virtual work. We have identified the entropy production on a curved surface which generalises the entropy production of bulk fluids known from irreversible thermodynamics to surfaces of arbitrary shapes \cite{de2013non}. Using this entropy production, we have identified conjugate fluxes and forces for an active fluid membrane. Our approach can also be directly applied to the study of active elastic surfaces. Our constitutive relations for active surfaces include the derivation of a fully generalized Hooke's law for elastic thin shells. We have classified active surfaces in 5 different symmetry classes: (i) up-down symmetric, non-chiral surfaces, (ii), non-chiral surfaces with broken up-down symmetry, (iii) chiral surfaces, (iv) planar chiral surfaces, and (v) chiral surfaces with broken up-down symmetry. Classes (i) and (ii) have been characterised before. Chiral surfaces (iii) must consist of chiral constituents and are up-down asymmetric, while planar-chiral surfaces (iv) do not have to be built from chiral subunits and only appear chiral when viewed from one side (Fig. \ref{fig:symmetries}). The constitutive equations for the surface have to obey these symmetries and coupling terms in the constitutive equation can be associated with these symmetry classes.

We have here neglected some degrees of freedom of the surface, such as the local rotation rate of molecules $\Omega_{ij}$ which relaxes rapidly to the vorticity flow $\omega_{ij}$ \cite{furthauer2012active}. We have also identified the normal derivative of the surface deformation with the rotation of the normal vector to the surface (Eqs. \ref{NormalDerivativeDisplacement} and \ref{CurlExpression}). This corresponds to neglecting a component of the shear normal to the surface. This additional contribution could be taken into account by adding an additional polar field tangential to the surface. We have considered the physics of an isolated surface, not taking into account the environment and external forces. Furthermore, we have restricted ourselves to isotropic surfaces. It will be interesting to expand the theory presented here to the case of active nematic or polar surfaces. 

When the surface is embedded in a viscous fluid, external forces and torques acting on the surface arise from stresses acting within the bulk fluid. The hydrodynamics of the 3D fluid and of the membrane are then coupled to each other. It would be interesting to expand the theory obtained here to include these couplings between the surface and the bulk fluid.

 Our work is related to previous works on active membrane and membrane dynamics  \cite{onuki1993dynamic, kralchevsky1994theory, ramaswamy2000nonequilibrium,chen2004internal,gov2004membrane,  lomholt2005general, lomholt2006descriptions, arroyo2009relaxation, rahimi2012shape} as well as on works on thin active films \cite{salbreux2009hydrodynamics,furthauer2013active}. We propose here a generic framework for active surfaces that captures many aspects of the physics discussed in earlier works. In addition, we identify new active terms associated with internal tensions and bending moments. In particular we show the existence of active torque terms that can induce curvature changes. 
 
 Our general approach provides a framework for the study of complex morphological changes of active surfaces in biology, for example during morphogenesis of an organism or the formation of complex cell shapes. We have introduced a limited number of phenomenological parameters which capture the generic effects of a large variety of molecular processes in cells and tissues. We expect in particular that biological processes such as tissue folding, invagination and twisting can be captured by our theory \cite{martin2010integration,taniguchi2011chirality}. Fold formation could occur through apical constriction \cite{sawyer2010apical}, which corresponds to the establishment of a difference in apical and basal surface tension in an epithelium, resulting in a gradient of active bending moment. Our theoretical framework provides a formalism to study how such gradients can result in tissue folding. By quantifying forces and deformations in tissues, the phenomenological parameters we introduce could be experimentally measured. Active tensions and bending moments could be related to the spatial and temporal distribution of force-generating elements such as myosin molecular motors in a tissue, as has been done to estimate active stresses distribution in the cell cortex \cite{mayer2010anisotropies,sedzinski2011polar,behrndt2012forces}.
 
 In general, biological systems have both elastic and viscous properties that could be captured by a viscoelastic generalisation of our theory. However, in many cases either elastic or viscous properties dominate: plant morphogenesis is often described as an active elastic medium, while long-time behaviour of tissue flows during animal morphogenesis can be captured by a viscous limit, on time scales where cells can rearrange their neighbours \cite{ranft2010fluidization,etournay2015interplay}. It will be a future challenge to find analytic and numerical solutions for the complex shape changes predicted by our theory. 

\acknowledgements

We thank Jacques Prost, Andrew Callan-Jones and Marino Arroyo for critical reading of the manuscript, and H\'el\`ene Berthoumieux, Karsten Kruse, Stephan Grill and Vijaykumar Krishnamurthy for useful discussions. G.S acknowledges support by the Francis Crick Institute which receives its core funding from Cancer Research UK (FC001317), the UK Medical Research Council (FC001317), and the Wellcome Trust (FC001317).





\appendix

\section{Differential geometry}
\label{AppendixDifferentialGeometry}

We give here definitions of differential geometry quantities used in the text. We consider a two-dimensional surface parametrized by two coordinates $\mathbf{X}(s^1,s^2)$.  Two tangent vectors and a normal vector are associated to every point on the surface, according to
\begin{eqnarray}
\mathbf{e}_1=\frac{\partial {\mathbf{X}}}{\partial s^1}\mbox{ , }\mathbf{e}_2=\frac{\partial {\mathbf{X}}}{\partial s^2}\mbox{ , }\mathbf{n}=\frac{\mathbf{e}_1\times\mathbf{e}_2}{|\mathbf{e}_1\times\mathbf{e}_2|}.
\end{eqnarray}
Lower indices correspond to covariant coordinates and upper indices to contravariant coordinates. The metric $g_{ij}$ and curvature tensor $C_i{}^j$ associated to $\mathbf{X}$ are defined by
\begin{eqnarray}
\label{MetricAndCurvature}
g_{ij}={ \mathbf e_i}.{ \mathbf e_j}\mbox{ , } C_{ij}=-(\partial_i\partial_j \mathbf{X}).\mathbf{n},
\end{eqnarray}
where $C_i{}^j=C_{ik}g^{kj}$. The inverse of the metric tensor $g^{ij}={g_{ij}}^{-1}$ verifies
\begin{equation}
g_{ij}g^{jk}=\delta_i{}^k.
\end{equation}
The contravariant basis is defined by
\begin{equation}
\mathbf{e}_i.\mathbf{e}^j=\delta_i{}^j,
\end{equation}
with $\mathbf{e}^i=g^{ij}\mathbf{e}_j$.
Indices can be raised and lowered by contraction with the metric tensor according to $a^i=g^{ij} a_j$ and $a_i=g_{ij}a^j$ for a tangent vector $\mathbf{a}=a^i\mathbf{e}_i=a_i \mathbf{e}^i$.

The derivatives of the basis and normal vectors are given by the Gauss-Weingarten equations
\begin{eqnarray}
\partial_i { \mathbf n}&=&C_i{}^j  { \mathbf e_j},\\
\partial_i \mathbf{e}_j&=&-C_{ij}\mathbf{n}+\Gamma_{ij}^k \mathbf{e}_k,
\end{eqnarray}
where the Christoffel symbols $\Gamma_{ij}^k$ are obtained from the metric by
\begin{equation}
\Gamma_{ij}^k=\frac{1}{2} g^{km}\left[\partial_j g_{im} + \partial_{i} g_{jm}-\partial_m g_{ij}\right].
\end{equation}
The surface area element is denoted $dS=\sqrt{g}ds^1 ds^2$ where $g=\det(g_{ij})$ is the determinant of the metric. 

The Levi-Civita tensor on the curved surface can be defined as:
\begin{equation}
\label{DefinitionEpsilonij}
\epsilon_{ij}=\mathbf{n}\cdot(\mathbf{e}_i\times\mathbf{e}_j).
\end{equation}
It is antisymmetric when expressed in a purely contravariant or covariant basis:
\begin{equation}
\label{DefinitionMatrixEpsilonij}
\epsilon_{ij}=\sqrt{g} \left(\begin{array}{cc}0 & 1 \\-1 & 0\end{array}\right)\mbox{,       }
\epsilon^{ij}=\frac{1}{\sqrt{g}} \left(\begin{array}{cc}0 & 1 \\-1 & 0\end{array}\right).
\end{equation}
Furthermore, it satisfies the identity
\begin{equation}
\epsilon_{ij}\epsilon^{jk}=-\delta_i^k.
\end{equation}
The Levi-Civita tensor can be used to express vectorial products of the basis vectors:
\begin{eqnarray}
\label{CrossProductIdentity1}
\textbf{n}\times\textbf{e}_i=\epsilon_i{}^{j} \textbf{e}_j ,\\
\label{CrossProductIdentity2}
\mathbf{e}_i\times\mathbf{e}_j=\epsilon_{ij}\mathbf{n}.
\end{eqnarray}
The second relation implies
\begin{equation}
|\mathbf{a}\times\mathbf{b}|=\epsilon_{ij}a^ib^j,
\end{equation}
for two tangent vectors $\mathbf{a}$ and $\mathbf{b}$.

A tensor with two indices can generally be decomposed into a symmetric and antisymmetric part:
\begin{eqnarray}
A_{ij}&=&A_{ij}^s+A_{ij}^a\label{TensorSymmetricAntisymmetricDecomposition}\\
&=&A_{ij}^s+\frac{1}{2} A_{kl}\epsilon^{kl} \epsilon_{ij}.
\end{eqnarray}

We denote $\partial_i=\partial/\partial{s^i}$ and $\nabla_i$ the covariant derivative, which has the property for a {\it tangent} vector $\mathbf{a}=a^i\mathbf{e}_i$ and tensor $\mathbf{t}=t^{ij} \mathbf{e}_i \otimes\mathbf{e}_j$:
\begin{eqnarray}
\nabla_i a^j&=&(\partial_i \mathbf{a})\cdot\mathbf{e}^j,\\
\nabla_i t^{jk}&=&\mathbf{e}^j\cdot(\partial_i \mathbf{t})\cdot\mathbf{e}^k,
\end{eqnarray}
 The definitions above then correspond to the following expressions:
\begin{eqnarray}
\nabla_i v^j&=&\partial_i v^j+\Gamma_{ik}^j v^k,\\
\nabla_i t^{jk}&=&\partial_i t^{jk}+\Gamma_{il}^j t^{lk}+\Gamma_{il}^k t^{jl}.
\end{eqnarray}
For a general vector $\mathbf{f}=f^i \mathbf{e}_i +f_n \mathbf{n}$, we have
\begin{equation}
\label{AppendixVectorDifferentation}
\partial_i \mathbf{f}=\left[\nabla_i f^j+ C_i{}^j f_n\right]\mathbf{e}_j+\left[\partial_i f_n-C_{ij}f^j\right]\mathbf{n}.
\end{equation}

The curvature tensor satisfies the Mainardi-Codazzi equation \cite{kreyszig1968introduction}:
\begin{equation}
\label{MainardiCoddazi}
\nabla_i C_{jk}=\nabla_j C_{ik}.
\end{equation}
The curvature tensor also satisfies the identity
\begin{eqnarray}
\nabla_i\left[C^i{}_j-C_k{}^k \delta^i{}_j\right]=0\label{DivergenceCurvatureTensor},
\end{eqnarray}
and the Gauss equation \cite{kreyszig1968introduction}
\begin{equation}
\label{DesernoGaussEquation}
C_{ik} C^{k}{}_j =C_k{}^k C_{ij}- g_{ij} \det(C_{k}{}^{l}).
\end{equation}

The covariant derivatives of the metric and of the Levi-Civita antisymmetric tensor vanish
\begin{eqnarray}
\nabla_i g^{jk}=0,\\
\nabla_i \epsilon^{jk}=0.
\end{eqnarray}

%
The coordinates of the tangent vectors in the 3D space with cartesian euclidian basis $\mathbf{u}_{\alpha}$ are written
\begin{eqnarray}
\mathbf{e}_i&=&e_{i,\alpha} \mathbf{u}_{\alpha}\\
\mathbf{e}^i&=&e^{i}_{\alpha} \mathbf{u}_{\alpha}.
\end{eqnarray}



The gradient of a vector field $\mathbf{v}(x_{\alpha})$ in the 3D space can be evaluated on the surface $\mathbf{X}$ through
\begin{equation}
\frac{\partial v_{\beta}}{\partial x_{\alpha}}=(\partial_i v_{\beta}) e^i_{\alpha}+ (\partial_n v_{\beta})n_{\alpha},
\end{equation}
where $\partial_n \mathbf{v}$ is the derivative normal to the surface. In particular, the curl of a vector field on the surface is given by
\begin{equation}
\label{AppendixCurlExpression}
(\boldsymbol{\nabla}\times\mathbf{v})_{\alpha}=\epsilon_{\alpha\beta\gamma} e^i_{\beta} \partial_i v_{\gamma} +\epsilon_{\alpha\beta\gamma} n_{\beta} \partial_n v_{\gamma}.
\end{equation}
The divergence theorem on a curved surface can be expressed using the covariant derivative \cite{capovilla2002stresses}:
\begin{equation}
\label{DivergenceTheorem}
\int_{\mathcal{S}} dS \nabla_i f^i=\int_{\mathcal{C}} dl \nu_i f^i,
\end{equation}
where $\mathcal{S}$ is the surface enclosed by $\mathcal{C}$, $\boldsymbol{\nu}$ is a unit vector tangent to $\mathcal{S}$, outward-pointing and normal to the contour $\mathcal{C}$, and $dl$ is an infinitesimal line element going along the contour $\mathcal{C}$. Eq. \ref{DivergenceTheorem} results from the identity \cite{kreyszig1968introduction}
\begin{equation}
\partial_i \sqrt{g}=\sqrt{g} \Gamma_{ki}^k .
\end{equation}
Indeed, denoting $s$ a coordinate going along the closed contour $\mathcal{C}$ in a trigonometric orientation around the normals to the surface $\mathcal{S}$, one obtains
\begin{eqnarray}
\int_{\mathcal{S}} dS \nabla_i f^i&=&\int_{\mathcal{S}} ds^1 ds^2 \partial_i (\sqrt{g} f^i)\nonumber\\
&=&\int_{\mathcal{C}} ds \sqrt{g}\left[\frac{\partial s^2}{\partial s}f^1-\frac{\partial s^1}{\partial s}f^2\right]\nonumber\\
&=&- \int_{\mathcal{C}} ds \frac{\partial s^i}{\partial s}\epsilon_{ij} f^j\nonumber\\
&=& \int_{\mathcal{C}} dl \nu_i f^i,
\end{eqnarray}
where the second line results from the usual divergence theorem, the third line from Eq. \ref{DefinitionMatrixEpsilonij}, and the fourth line from the relations
\begin{eqnarray}
dl&=&ds |\mathbf{e}_s|,\\
\boldsymbol{\nu}&=&\frac{\mathbf{e}_s\times\mathbf{n}}{|\mathbf{e}_s\times\mathbf{n}|}=-\frac{\partial s^i}{\partial s} \epsilon_i{}^j \mathbf{e}_j/|\mathbf{e}_s|,
\end{eqnarray}
with $\mathbf{e}_s=\partial_s \mathbf{X}=(\partial s^i/\partial s) \mathbf{e}_i$ the vector tangent to the contour $\mathcal{C}$.

\section{Variation of surface quantities}
\label{AppendixVariationSurfaceQuantities}
We consider here that the surface $\mathbf{X}$ is modified to a new surface $\mathbf{X}'$:
\begin{equation}
\mathbf{X'}(s^1,s^2)=\mathbf{X}(s^1,s^2)+\mathbf{\delta X}(s^1,s^2).
\end{equation}
We derive here expressions for the perturbations of the associated differential geometry quantities. The tangent vector variation reads
\begin{equation}
\delta \mathbf{e}_i=\partial_i \delta \mathbf{X}.
\end{equation}
Using $g_{ij}=\partial_i \mathbf{X}\cdot\partial_j \mathbf{X}$, one finds
\begin{equation}
\label{Expressiondeltagij}
\delta g_{ij}=(\partial_i \delta\mathbf{X})\cdot\mathbf{e}_j+(\partial_j \delta\mathbf{X})\cdot\mathbf{e}_i.
\end{equation}
Using $\mathbf{n}\cdot\mathbf{e}_i=0$ and $\mathbf{n}\cdot\mathbf{n}=0$,
\begin{equation}
\label{VariationNormalVectorFirstExpression}
\delta \mathbf{n}=-((\partial_i \delta \mathbf{X})\cdot\mathbf{n}) \mathbf{e}^i.
\end{equation}
Using $\mathbf{e}^i\cdot \mathbf{e}_j=\delta^i_j$, resulting in $\mathbf{e}^i\cdot\delta\mathbf{e}_j+\delta\mathbf{e}^i\cdot\mathbf{e}_j=0$,
\begin{equation}
\delta \mathbf{e}^i=-((\partial_j \delta \mathbf{X})\cdot\mathbf{e}^i)\mathbf{e}^j+((\partial^i \delta \mathbf{X})\cdot\mathbf{n}) \mathbf{n}.
\end{equation}
Using $C_{ij}=-(\partial_i \partial_j \mathbf{X})\cdot\mathbf{n}$ and $C_i{}^j=C_{ik}g^{kj}$,
\begin{eqnarray}
\label{ExpressionDeltaCij}
\delta \hat{C}_{ij}
&=&-(\nabla_i \partial_j \mathbf{\delta X})\cdot\mathbf{n},\\
\label{ExpressionDeltaCijMixed}
\delta C_i{}^j&=&\delta \hat{C}_{ik} g^{kj} + C_{ik} \delta g^{kj}.
\end{eqnarray}
Note that we distinguish $\delta\hat{C}_{ij}=C'_{ij}-C_{ij}$ and $\delta C_i{}^j=C'_i{}^j-C_i{}^j$, which are two different tensors, related by Eq. \ref{ExpressionDeltaCijMixed}.

Using $\Gamma_{ij}^k=(\partial_i\partial_j \mathbf{X})\cdot\mathbf{e}^k$,
\begin{eqnarray}
\label{ExpressionDeltaGammaijk}
\delta\Gamma_{ij}^k
&=&(\nabla_i\partial_j \delta \mathbf{X})\cdot\mathbf{e}^k-C_{ij} (\partial^k \delta \mathbf{X})\cdot\mathbf{n}
\end{eqnarray}
Separating $\mathbf{\delta X}$ into a tangent and normal part:
\begin{equation}
\mathbf{\delta X}=\delta X^i \mathbf{e}_i+\delta X_n \mathbf{n},
\end{equation}
we obtain the expressions in terms of components of the shape perturbation:
\begin{eqnarray}
\delta\mathbf{e}_i&=&(\nabla_i \delta X^j + C_i{}^j \delta X_n ) \mathbf{e}_j+(\partial_i  \delta X_n- C_{ij}\delta X^j  ) \mathbf{n},\label{VariationTangentVector}\\
\delta\mathbf{e}^i&=&-(\nabla^j\delta X^i+C^{ij} \delta X_n )\mathbf{e}_j+(\partial^i \delta X_n-C^i{}_j\delta X^j )\mathbf{n},\\
\delta\mathbf{n}&=&(-\partial_i  \delta X_n+C_{ij}\delta X^j  ) \mathbf{e}^i,\label{VariationNormalVector}\\
\delta g_{ij}&=&\nabla_i \delta X_j+\nabla_j \delta X_i+2C_{ij} \delta X_n ,\label{VariationMetricCovariant}\\
\delta g^{ij}&=&-\nabla^i \delta X^j -\nabla^j \delta X^i -2C^{ij}\delta X_n, \label{VariationMetricContravariant} \\
\delta \sqrt{g}&=&\frac{1}{2}\sqrt{g} g^{ij}\delta g_{ij},\label{VariationSqrtDeterminantMetric}\\
\delta \hat{C}_{ij}&=&-\nabla_i (\partial_j \delta X_n)+(\nabla_j \delta X^k) C_{ik}+(\nabla_i \delta X^k) C_{kj}+(\nabla_i C_{jk})\delta X^k+\delta X_nC_{ik} C^{k}{}_{j}, \\
\delta C_i{}^j&=&-\nabla_i (\partial^j \delta X_n)+(\nabla_i \delta X^k) C_{k}{}^{j}-(\nabla^k \delta X^j)C_{ik} +(\nabla_i C^{j}{}_{k})\delta X^k-  \delta X_n C_{ik} C^{kj},\label{VariationCurvatureTensor}\\
\delta \Gamma_{ij}^k&=&\nabla_i \nabla_j \delta X^k +(C_{ij} C^k{}_l-C_{jl}C_i{}^k)\delta X^l+ C_j{}^k (\partial_i \delta X_n) +C_i{}^k (\partial_j \delta X_n)  -C_{ij} (\partial^k \delta X_n) \nonumber\\
&& +\nabla_i C_j{}^k \delta X_n \label{VariationChristoffel}.
\end{eqnarray}


In order to define the  the normal derivative of an infinitesimal surface deformation, $\partial_n \delta \mathbf{X}$, we introduce material coordinates for the points in the volume around the surface:
\begin{eqnarray}
\mathbf{\overline{X}}(s^1,s^2,z)=\mathbf{X}(s^1,s^2)+z \mathbf{n},
\end{eqnarray}
with $z$ a coordinate going along the normal to the surface. When the surface is deformed with infinitesimal vector deformation $\delta\mathbf{X}$, we assume that the volume around the surface is deformed by
\begin{eqnarray}
\delta \mathbf{\overline{X}}(s^1,s^2,z)=\delta \mathbf{X}(s^1,s^2)+z \delta \mathbf{n}.
\end{eqnarray}
This choice implies that only in-plane shear occurs. We then obtain
\begin{equation}
\label{NormalDerivativeDisplacement}
\partial_z \delta\mathbf{\overline{X}}=\delta \mathbf{n}=-(\partial_i \delta \mathbf{X}\cdot\mathbf{n})\mathbf{e}^i.
\end{equation}
We identify $\partial_n \delta \mathbf{X}$ with $\partial_z \delta \mathbf{\overline{X}}$ in Eq. \ref{CurlExpression}. This choice is equivalent to assume that points along the normal to the initial surface before deformation are along the normal to the new surface after deformation.

\section{Force balance derivation}
\label{ForceBalanceDerivation}
We discuss here the force and torque balance for an element of surface. We consider a force balance equation taking into account the contribution of mass accretion or ejection from the surface. For simplicity, we assume here that mass accretion or ejection occurs only on one side of the surface. Applying the law of Newton on a surface region $\mathcal{S}$ of contour $\mathcal{C}$ yields
\begin{eqnarray}
\label{InitialForceBalanceEquation}
\partial_t\left(\int_\mathcal{S}dS \rho \mathbf{v}\right)&=&\int_\mathcal{S}  dS J_n^{\rho}(\mathbf{v}+\mathbf{u})+\oint_\mathcal{C} dl \nu_i \mathbf{t}^i+\int_\mathcal{S} dS {\mathbf{f}}_0^{\rm{ext}},
\end{eqnarray}
where the second term arises from the change of momentum due to mass being absorbed by the surface with velocity $\mathbf{u}$ relative to the surface, and the third term arises from the force acting on the surface $S$ from the surface outside of $S$. $\mathbf{f}_0^{\rm{ext}}$ is the external stress acting on the surface in addition to the momentum of incoming molecules. The flux of mass towards the surface $J_n^{\rho}$ is introduced in Eq. \ref{MassBalanceEquation}. The surface momentum rate of change can be rewritten using Eqs. \ref{TimeDerivativeSqrtDetMetricEulerian}, \ref{MassBalanceEquation}, and the divergence theorem \ref{DivergenceTheorem}:
\begin{eqnarray}
\partial_t\left(\int_\mathcal{S} dS \rho \mathbf{v}\right)&=&\int_\mathcal{S} dS \left[\partial_t(\rho \mathbf{v})+v_n C_k{}^k \rho \mathbf{v}\right]+\oint_\mathcal{C} d \nu_i v^i \rho \mathbf{v}\nonumber \\
&=&\int_\mathcal{S} dS \bigg[\rho \partial_t \mathbf{v}- \nabla_i (\rho v^i)\mathbf{v} +J_n^{\rho} \mathbf{v}+\nabla_i(\rho v^i \mathbf{v}) \bigg]\nonumber\\
&=&\int_\mathcal{S} dS \bigg[\rho ( \partial_t \mathbf{v}+ v^i \nabla_i \mathbf{v} )+J_n^{\rho} \mathbf{v} \bigg],
\end{eqnarray}
such that the force balance equation \ref{InitialForceBalanceEquation} can be rewritten
\begin{equation}
\label{InitialForceBalance}
\int_\mathcal{S} dS \rho \mathbf{a}=\oint_\mathcal{C} dl \nu_i \mathbf{t}^i+\int_\mathcal{S} dS \mathbf{f}^{\rm{\rm{ext}}},
\end{equation}
where we have introduced the total external force $\mathbf{f}^{\rm{ext}}=\mathbf{f}_0^{\rm{ext}}+J_n^{\rho} \mathbf{u}$ , and the acceleration $\mathbf{a}$ is defined by
\begin{equation}
\label{AccelerationDefinition}
\mathbf{a}=\frac{d \mathbf{v}}{dt}
\end{equation}
with $d/dt=\partial_t + v^i \nabla_i$ the convected derivative. Using the divergence theorem \ref{DivergenceTheorem}, Eq. \ref{InitialForceBalance} can be rewritten
\begin{equation}
\int_\mathcal{S} dS \left[\rho \mathbf{a}-\nabla_i \mathbf{t}^i - \mathbf{f}^{\rm{\rm{ext}}}\right]=0.
\end{equation}
Because this equation has to be valid for any surface element, this results in Eq. \ref{ForceBalanceEquation}, which can also be written in the form of a local conservation of momentum:
\begin{equation}
\partial_t (\rho \mathbf{v})=\nabla_i ( \mathbf{t}^i -\rho v^i \mathbf{v}) +\mathbf{f}_0^{\rm{ext}} - \rho C_k{}^k v_n \mathbf{v}+J_n^\rho(\mathbf{v}+\mathbf{u}).
\end{equation}
Here the three last terms arise from exchange of momentum normal to the surface.

Ignoring the moment of inertia tensor for simplicity, the total torque acting on on a surface region $\mathcal{S}$ of contour $\mathcal{C}$ vanishes:
\begin{eqnarray}
\oint_\mathcal{C} dl \nu_i \mathbf{m}^i+\oint_\mathcal{C} dl \mathbf{X}\times\nu_i\mathbf{t}^i+\int_\mathcal{S} dS \mathbf{X}\times (\mathbf{f}^{\rm{ext}}-\rho \mathbf{a})+\int_\mathcal{S} dS \boldsymbol{\Gamma}^{\rm{ext}}=0,
\end{eqnarray}
where $\boldsymbol{\Gamma}^{\rm{ext}}$ is the external torque density acting on the surface. Using the divergence theorem and the force balance equation \ref{ForceBalanceEquation}, the torque balance equation can be rewritten
\begin{eqnarray}
\int_\mathcal{S} dS \left[\nabla_i \mathbf{m}^i+ \mathbf{e}_i \times\mathbf{t}^i+ \boldsymbol{\Gamma}^{\rm{ext}}\right]=0
\end{eqnarray}
which results in the torque balance expression Eq. \ref{TorqueBalanceEquation}.

We note that the force balance equations \ref{ForceBalanceTangential}-\ref{TorqueBalanceNormal} are invariant under the variable transformation
\begin{eqnarray}
\label{GaugeTransformationtij}
t^{ij}&\rightarrow& t^{ij}+ m  \epsilon^i{}_k C^{kj},\\
t_n^i&\rightarrow&t_n^i+ \epsilon^{ij} (\partial_j m),\\
\label{GaugeTransformationmij}
m^{ij}&\rightarrow& m^{ij} +m g^{ij}.
\end{eqnarray}
where $m$ is an arbitrary function on the surface, and we have used equations \ref{MainardiCoddazi} and \ref{DesernoGaussEquation}. 

\section{Differential work}
\label{DifferentialWork}

The virtual work defined in Eq. \ref{VirtualWorkDefinition} can be rewritten using the divergence theorem on a curved surface \ref{DivergenceTheorem} and the force and torque balance equations \ref{ForceBalanceEquation} and \ref{TorqueBalanceEquation}:
\begin{eqnarray}
\delta W&=&\int_\mathcal{S} dS\bigg[ \mathbf{t}^i\cdot \partial_i \mathbf{\delta X}+\frac{1}{2} \mathbf{m}^{i}\cdot\partial_i (\boldsymbol{\nabla}\times \delta\mathbf{X})+\frac{1}{2} ( \mathbf{t}^i\times \mathbf{e}_i )\cdot(\boldsymbol{\nabla}\times \delta\mathbf{X})\bigg].
\end{eqnarray}
Projecting $t^i$ and $m^i$ along the tangent and normal directions and using Eqs. \ref{CrossProductIdentity1}-\ref{CrossProductIdentity2}, one finds
\begin{eqnarray}
\delta W&=&\int_\mathcal{S} dS\bigg[ t^{ij}\mathbf{e}_j\cdot \partial_i \mathbf{\delta X}+t^{i}_n \mathbf{n}\cdot \partial_i \mathbf{\delta X}+\frac{1}{2} m^{ij} \mathbf{e}_j\cdot\partial_i (\boldsymbol{\nabla}\times \delta\mathbf{X})+\frac{1}{2} m^{i}_n \mathbf{n}\cdot \partial_i (\boldsymbol{\nabla}\times \delta\mathbf{X})\nonumber\\
&&-\frac{1}{2} t^{ij} \epsilon_{ij} \mathbf{n}\cdot(\boldsymbol{\nabla}\times\delta \mathbf{X})+\frac{1}{2} t^{i}_{n} \epsilon_{ij}\mathbf{e}^j\cdot (\boldsymbol{\nabla}\times \delta\mathbf{X})\bigg].
\end{eqnarray}

Using the definition of the curl operator Eq. \ref{CurlExpression}, the relations \ref{CrossProductIdentity1} and \ref{CrossProductIdentity2}, the expression of the normal derivative of the displacement \ref{NormalDerivativeDisplacement}, the variation of the curvature tensor \ref{ExpressionDeltaCij} and of the Christoffel symbols \ref{ExpressionDeltaGammaijk}, the following identities can be obtained:
\begin{eqnarray}
\mathbf{n}\cdot(\boldsymbol{\nabla}\times\delta\mathbf{X})
&=& \epsilon^{i}{}_{j} \partial_i \delta \mathbf{X} \cdot \mathbf{e}^j\\
\mathbf{e}^i\cdot(\boldsymbol{\nabla}\times\delta\mathbf{X})
&=&2\epsilon^{ij} \partial_j \delta \mathbf{X} \cdot \mathbf{n}\label{TangentialCurlInfinitesimalDisplacement}\\
\mathbf{e}_j\cdot\partial_i (\boldsymbol{\nabla}\times \delta\mathbf{X})
&=&\left[2C_i{}^l \epsilon_j{}^k +C_{ij}\epsilon^{kl}\right]\partial_k \delta\mathbf{X}\cdot\mathbf{e}_l-2\epsilon_j{}^k \delta \hat{C}_{ik}\\
\mathbf{n}\cdot\partial_i (\boldsymbol{\nabla}\times \delta\mathbf{X})&=&\epsilon^{j}{}_{k} \delta \Gamma_{ij}^k.
\end{eqnarray}
Using these relations, the virtual work can be rewritten
\begin{eqnarray}
\delta W&=&\int_\mathcal{S} dS\bigg[\bigg(t^{ij}-\frac{1}{2} t^{kl} \epsilon_{kl} \epsilon^{ij} +m^{kl} C_k{}^j \epsilon_l{}^i + \frac{1}{2}m^{kl}C_{kl}\epsilon^{ij}\bigg) \partial_i \mathbf{\delta X}\cdot\mathbf{e}_j-m^{ij} \epsilon_j{}^k \delta \hat{C}_{ik}+\frac{1}{2} m^{i}_n \epsilon^j{}_k \delta \Gamma_{ij}^k \bigg].
\end{eqnarray}
Using the in-plane torque tensor introduced in Eq. \ref{DefinitionInPlaneTorqueTensor} $\bar{m}^{ij}=-m^{ik}\epsilon_k{}^j$ (with inverse relation $m^{ij}=\bar{m}^{ik}\epsilon_k{}^j$) and using the expression for the variation of the metric \ref{Expressiondeltagij}, one finds
\begin{eqnarray}
\delta W&=&\int_\mathcal{S} dS\bigg[\frac{1}{2} (t^{ij}_s-\frac{1}{2}(\bar{m}^{ki} C_k{}^j+\bar{m}^{kj} C_k{}^i)) \delta g_{ij}+\bar{m}^{ij}  \delta \hat{C}_{ij}+\frac{1}{2} m^{i}_n \epsilon^j{}_k \delta \Gamma_{ij}^k \bigg].
\end{eqnarray}

Using Eq. \ref{ExpressionDeltaCijMixed} leads to the alternative expression of the virtual work
\begin{eqnarray}
\delta W&=&\int_S dS\bigg[\frac{1}{2} (t^{ij}_s+\frac{1}{2}(\bar{m}^{ki} C_k{}^j+\bar{m}^{kj} C_k{}^i)) \delta g_{ij}+\bar{m}^{i}{}_{j}  \delta C_{i}{}^{j}+\frac{1}{2} m^{i}_n \epsilon^j{}_k \delta \Gamma_{ij}^k \bigg],
\end{eqnarray}
which leads to Eq. \ref{VirtualWork2} with  the definition \ref{DefinitionInPlaneStressTensor}.

The deformation term in factor of $m_n^i$ is a generalisation to curved surface of the gradient of rotations \cite{furthauer2012active}. This can be seen from its explicit expression in term of the deformation coordinates:
\begin{equation}
\epsilon^j{}_k \delta \Gamma_{ij}^k=\nabla_i (\epsilon^j{}_k \nabla_j \delta X^k)-2C_{ij} \epsilon^{jk} (\partial_k \delta X_n -C_{kl} \delta X^l),\label{GradientRotationTensor}
\end{equation}
where we have used Eq. \ref{VariationChristoffel}.

\section{Eulerian and Lagrangian representation of surface flows}
\label{EulerianLagrangian}
\subsection{Lagrangian representation}

In a Lagrangian representation, the parameters $s^1$ and $s^2$ label the center of mass of a specific volume element.
The surface is characterised by the time-dependent parametrisation $\mathbf{X}(s^1,s^2,t)$. The center-of-mass velocity is given by 
\begin{equation}
\mathbf{v}=\partial_t \mathbf{X}(s^1,s^2,t).
\end{equation}

The mass density conservation equation without exchange between the surface and its environment reads in Lagrangian coordinates
\begin{equation}
\label{AppendixLagrangianMassConservation}
\partial_t \rho +\rho(\nabla_i v^i +C_i{}^i v_n)=0.
\end{equation}
This can be seen from the conservation of mass of a region of surface $\mathcal{S}$:
\begin{eqnarray}
\frac{d}{dt}\left( \int_\mathcal{S} dS \rho(s^1,s^2)\right)=0\\
\int_\mathcal{S} ds^1ds^2\left(\rho \partial_t (\sqrt{g})+\sqrt{g} \partial_t \rho \right)=0\\
\int_\mathcal{S} dS \left(\left(\nabla_i v^i+v_n C_i{}^i\right)\rho +\partial_t \rho \right)=0.
\end{eqnarray}
which leads to Eq. \ref{AppendixLagrangianMassConservation}.  In this derivation, we have obtained  $d\sqrt{g}/dt$ by setting $\delta \mathbf{X}=\mathbf{v} dt$ in Eq. \ref{VariationSqrtDeterminantMetric}.

The rate of change of metric in Lagrangian coordinates
\begin{equation}
\label{LagrangianDerivativeMetric}
\partial_t g_{ij}=\nabla_i v_j+\nabla_j v_i + 2 v_n C_{ij}
\end{equation}
obtained by setting $\delta \mathbf{X}=\mathbf{v} dt$ in Eq. \ref{VariationMetricCovariant}, relates to the gradient of flow defined in Eq. \ref{DefinitionSymmetricVelocityGradient} through $\partial_t g_{ij}=2 v_{ij}$.

Similarly, the rate of change of the curvature tensor in Lagrangian coordinates, obtained by setting $\delta \mathbf{X}=\mathbf{v} dt$ in Eq. \ref{VariationCurvatureTensor}, defines a convected Lagrangian derivative $\bar{D}/{\bar{D}t}$ of the curvature tensor:
\begin{eqnarray}
\label{LagrangianDerivativeCurvatureTensor}
\frac{\bar{D}C_i{}^j}{\bar{D}t}&=&-\nabla_i(\partial^j v_n)
-v_n C_{ik} C^{kj} +(\nabla_i C^j{}_k) v^k +(\nabla_i v^k) C_k{}^j -(\nabla_k v^j) C_{i}{}^{k}
\end{eqnarray}
and its symmetric part is introduced in Eq. \ref{DefinitionCorotationalCurvatureTensor}. The rate of change of the Christoffel symbols is related to the gradient of rotations introduced in Eqs. \ref{RateDissipationFreeEnergy} and \ref{DefinitionFlowRotational} through the identity
\begin{equation}
\label{LagrangianDerivativeChristoffel}
\frac{1}{2} \epsilon^j{}_k \frac{\bar{D}\Gamma_{ij}^k}{\bar{D}t}=\nabla_i \omega_n-C_{ij} \omega^j=(\partial_i \boldsymbol{\omega})\cdot\mathbf{n},
\end{equation}
where we have used Eq. \ref{VariationChristoffel}.
\subsection{Eulerian coordinates}

In Eulerian coordinates, the center-of-mass velocity field is given by
\begin{equation}
\mathbf{v}=v^i\mathbf{e}_i+v_n\mathbf{n},
\end{equation}
where $v^i\mathbf{e}_i$ is the tangential velocity field, and the normal velocity field is given by
\begin{equation}
v_n=(\partial_t \mathbf{X}(s^1,s^2,t))\cdot\mathbf{n}.
\end{equation}
 In addition, one requires the condition
\begin{equation}
(\partial_t \mathbf{X}(s^1,s^2,t))\cdot\mathbf{e}^i=0.
\end{equation}
such that coordinates do not change when the surface is not deforming. Here, $s^1$ and $s^2$ do not describe a specific material element.

In the Eulerian perspective, the time derivative of the tangent vectors, normal, metric, surface element area and curvature are given by
\begin{eqnarray}
\partial_t \mathbf{e}_i&=&v_n C_i{}^j \mathbf{e}_j+(\partial_i v_n)\mathbf{n}\label{TimeDerivativeTangentVectorEulerian}\\
\partial_t \mathbf{n}&=&-(\partial_i v_n) \mathbf{e}^i\label{TimeDerivativeNormalVectorEulerian}\\
\partial_t g_{ij}&=&2 v_n C_{ij}\label{TimeDerivativeMetricEulerian}\\
\partial_t \sqrt{g}&=&\sqrt{g}v_n C_i{}^i\label{TimeDerivativeSqrtDetMetricEulerian}\\
\partial_t C_{i}{}^{j}&=&-\nabla_i (\partial^j v_n)-v_n C_{ik}C^{kj}\label{TimeDerivativeCurvatureEulerian}
\end{eqnarray}
where we have used Eqs. \ref{VariationTangentVector}, \ref{VariationNormalVector}, \ref{VariationMetricCovariant}, \ref{VariationSqrtDeterminantMetric} and \ref{VariationCurvatureTensor} with $\delta X^i=0$ and $\delta X_n=v_n dt$.

Mass conservation without exchange between the surface and its environment has the form
\begin{equation}
\label{AppendixEulerianMassConservation}
\partial_t \rho +\nabla_i (\rho v^i) +\rho C_i{}^i v_n  =0,
\end{equation}
which follows from the mass conservation of an element of surface $\mathcal{S}$ with fixed contour $\mathcal{C}$:
\begin{eqnarray}
\frac{d}{dt}\left( \int_\mathcal{S} dS \rho\right)=-\oint_\mathcal{C} dl \boldsymbol{\nu}.\mathbf{v} \rho=-\int_\mathcal{S} dS \nabla_i (v^i \rho)\nonumber\\
\int_\mathcal{S} ds^1ds^2\left(( \partial_t \sqrt{g}) \rho +\sqrt{g} \partial_t \rho\right)=-\int_\mathcal{S} dS \nabla_i (v^i \rho)\nonumber\\
\int_\mathcal{S} dS \left(\left(v_n C_i{}^i\right)\rho +\partial_t \rho+\nabla_i (\rho v^i)\right)=0.
\end{eqnarray}
which leads to Eq. \ref{AppendixEulerianMassConservation}.

The acceleration reads (Eq. \ref{AccelerationDefinition}),
\begin{eqnarray}
\mathbf{a}&=&\partial_t \mathbf{v}+v^i\partial_i \mathbf{v}\\
&=&\left[ \partial_t v^i+v^j \nabla_j v^i+2 v_n v^j C_j{}^i -v_n \partial^i v_n\right]\mathbf{e}_i+\left[\partial_t v_n +2 v^i\partial_i v_n -v^i v^j C_{ij} \right]\mathbf{n}\label{DetailedEulerianExpressionAcceleration},
\end{eqnarray}
where we have used Eqs. \ref{TimeDerivativeTangentVectorEulerian} and \ref{TimeDerivativeNormalVectorEulerian}.
\section{Translation and rotation invariance}
\label{AppendixGibbsDuhem}

We derive here relations for the stress and torque tensor of a fluid surface, obtained from the invariance of the free energy under rigid translation and rotations of the surface. We consider for this derivation a surface in the absence of external forces. The fluid surface contains $N$ species $\alpha=1...N$ with concentration $c^{\alpha}$ and its free energy density is given by Eq. \ref{TotalFreeEnergyDensity}. The deformation by an infinitesimal rigid translation or rotation defines a new surface $\mathbf{X}'=\mathbf{X}+\delta\mathbf{X}$. The new surface is then reparameterized by new coordinates, such that a point $(s^1,s^2)$ on the initial surface finds its new position on the new surface by going along the normal to the initial surface (Figure \ref{fig:schematicGibbsDuhem}):
\begin{equation}
\label{AppendixNormalDisplacement}
\mathbf{X}''(s^1,s^2)=\mathbf{X}(s^1,s^2)+\delta \mathbf{X}\cdot\mathbf{n}\quad .
\end{equation}

\subsection{Invariance by translation}

We consider here a rigid translation of the surface by an infinitesimal uniform vector $\delta \mathbf{a}$, implying $\partial_i \delta\mathbf{a}=0$ and the relations \ref{RigidTranslationGibbsDuhem1}-\ref{RigidTranslationGibbsDuhem2}.
With the choice of coordinates \ref{AppendixNormalDisplacement}, the concentration, density and velocity fields on the surface are modified only by the tangential contributions of displacement:
 \begin{eqnarray}
 \delta c^{\alpha}&=&-\partial_i c^{\alpha} \delta a^i,\\
 \frac{1}{2}\delta(\rho v ^2)&=&-\frac{1}{2}\partial_i (\rho v^2)\delta a^i
 \end{eqnarray}
 The geometric quantities on the new surface can be obtained by using Eqs. \ref{VariationSqrtDeterminantMetric} and \ref{VariationCurvatureTensor}, with the normal displacement \ref{AppendixNormalDisplacement}, and using Eqs. \ref{RigidTranslationGibbsDuhem1} and \ref{RigidTranslationGibbsDuhem2}: 
\begin{eqnarray}
 \delta \sqrt{g}&=&\sqrt{g} C_i{}^i \delta a_n\\
 \delta C_{j}{}^{k}&=&-\nabla_i C_{j}{}^{k} \delta a^i
 \end{eqnarray}
The variation of surface free energy after the rigid translation must vanish, and is given by
\begin{eqnarray}
\delta F&=&\int_\mathcal{S} dS \left[f \delta a_n C_k{}^k +\delta f \right]+ \oint_\mathcal{C} d \nu_i (f \delta a^i)\nonumber\\
&=&\int_\mathcal{S} dS \bigg[ f \delta a_n C_k{}^k -\mu^{\alpha} \partial_i c^{\alpha} \delta a^i-\frac{1}{2}\partial_i(\rho v^2) \delta a^i-K^{jk} \nabla_i C_{jk}\delta a^i+\nabla_i(f \delta a^i) \bigg] \nonumber\\
&=&\delta a^i  \int_\mathcal{S} dS \left[ - (\mu^{\alpha} \partial_i c^{\alpha}+K^{jk}\nabla_i C_{jk})  +\partial_i f_0 \right] \nonumber\\
&=&\delta a^i  \int_\mathcal{S} dS \bigg[ \partial_i \mu^{\alpha} c^{\alpha} +(\nabla_j K^{jk} ) C_{ik}+\nabla_j( (f_0-\mu^{\alpha} c^{\alpha} )g_{i}{}^{j}- K^{jk} C_{ik}) \bigg]\nonumber \\
&=&0,\label{AppendixTranslationInvarianceGibbsDuhem}
\end{eqnarray}
where we have used the expression of the differential of the free energy density \ref{TotalFreeEnergyDensity} at constant temperature, Eq. \ref{RigidTranslationGibbsDuhem1}, and the Mainardi-Coddazi equation \ref{MainardiCoddazi}.

 Because Eq. \ref{AppendixTranslationInvarianceGibbsDuhem} is valid for any surface element and any infinitesimal vector $\delta\mathbf{a}$, we obtain Eq.\ref{GibbsDuhemTranslation}.

\subsection{Invariance by rotation}
We now consider a uniform rotation of the surface with vector $\delta\boldsymbol{ \theta}$, such that the surface is deformed by $\delta \mathbf{X}=\delta\boldsymbol{\theta}\times\mathbf{X}$. One can verify that the following identity holds for such a deformation:
\begin{equation}
\nabla_i \delta X^i=-C_i{}^i \delta X_n.\label{GibbsDuhemRotationIdentity}
\end{equation}
As for a rigid translation, the concentration, density and velocity fields on the surface are modified only by tangential contributions of displacements. One finds then
 \begin{eqnarray}
 \delta c&=&-(\partial_i c^{\alpha}) \delta X^i\\
 \frac{1}{2}\delta(\rho v^2)&=&-\frac{1}{2}\left(\partial_i(\rho v^2)\right) \delta X^i
 \end{eqnarray}
 As for translations, changes in geometric quantities can be obtained from Eqs. \ref{VariationSqrtDeterminantMetric} and \ref{VariationCurvatureTensor}, with the normal displacement \ref{AppendixNormalDisplacement}:
 \begin{eqnarray}
 \delta C_{i}{}^{j}&=&-(\nabla_k C_{i}{}^{j} )\delta X^k-\delta \omega^{jk} C_{ki}-\delta \omega_{ik} C^{kj} \label{RigidRotationCurvatureChange}\\
\delta(\sqrt{g})&=&\sqrt{g} C_i{}^i \delta X_n
\end{eqnarray}
 with $\delta \omega_{ij}=\epsilon_{ij} (\frac{1}{2}\nabla_k\delta X_l \epsilon^{kl})=\epsilon_{ij}\delta\boldsymbol{\theta}\cdot\mathbf{n}$. The associated variation of free energy reads
 \begin{eqnarray}
\delta F&=&\int_\mathcal{S} dS \left[f \delta X_n C_k{}^k +\delta f\right] + \oint_\mathcal{C} d \nu_i (f \delta X^i)\nonumber\\
&=&\int_\mathcal{S} dS \bigg[f \delta X_n C_k{}^k  -\mu^{\alpha} (\partial_i c^{\alpha}) \delta X^i-\frac{1}{2}\left(\partial_i(\rho v^2)\right)\delta X^i-K^{ij} (\nabla_k C_{ij})\delta X^k \\
&&-K^{ij} \delta \omega_{jk} C_{i}{}^{k}- K^{ij}\delta \omega_{ik} C^k{}_{j}+\nabla_i(f \delta X^i)\bigg]\nonumber\\ 
&=&\int_\mathcal{S} dS \bigg[ -(\mu^{\alpha} \partial_i c^{\alpha}+K^{jk}\nabla_i C_{jk}) \delta X^i +(\partial_i f_0) \delta X^i  -\delta\boldsymbol{\theta}\cdot\mathbf{n}\left[K^{ij} \epsilon_{jk} C_{i}{}^{k}+K^{ij}\epsilon_{ik} C^k{}_{j}\right]\bigg]\nonumber\\
&=& -\delta\boldsymbol{\theta}\cdot\mathbf{n}\int_S dS \left[K^{ij} \epsilon_{jk} C_{i}{}^{k}+K^{ij}\epsilon_{ik} C^k{}_{j}\right]\nonumber\\
&=& -2 \delta\boldsymbol{\theta}\cdot\mathbf{n}\int_S dS \epsilon_{jk} K^{ij} C_{i}{}^{k}\nonumber\\
&=&0,\label{AppendixGibbsDuhemRotation}
\end{eqnarray}
where we have used Eqs. \ref{GibbsDuhemTranslation} and \ref{GibbsDuhemRotationIdentity}.

 Because Eq. \ref{AppendixGibbsDuhemRotation} is valid for any surface element and any infinitesimal rotation vector $\delta\boldsymbol{\theta}$, this results in Eq. \ref{GibbsDuhemRotation}.
 
 \section{Up-down asymmetry, chirality and planar-chirality of surfaces}
 \label{AppendixSurfaceSymmetries}
 
 We discuss here the symmetries of a surface with rotational symmetry in the plane. The symmetries that can be broken for to these surfaces are the up-down mirror symmetry ($M_n$), the mirror symmetries in the plane ($M_t$, a mirror symmetry by a plane going along an arbitrary tangent vector $\mathbf{t}$), and the up-down rotation symmetries ($R_t$, a rotation of $\pi$ around an arbitrary tangent vector $\mathbf{t}$). Rotations around the normal with angle $\pi$, $R_{n}$ preserve the state of a surface with rotational symmetry in the plane. Composition of these symmetries are indicated in the multiplication table \ref{SymmetryGroupTable}. 
 
 \begin{table}[h!]
\centering
  \begin{tabular}{|c|c|c|c|c|}\hline
&$M_n$& $M_t$ &$R_t$ &$R_n$ \\\hline
$M_n$ & $\mathbb{1}$ &$R_t$ &$M_t$&$I$\\\hline
$M_t$ &$R_t$ &$\mathbb{1}$&$M_n$&$M_t R_n$ \\\hline
$R_t$ &$M_t$&$M_n$&$\mathbb{1}$&$R_t R_n$\\\hline
$R_n$ &$I$&$M_t R_n$&$R_t R_n$&$\mathbb{1}$ \\\hline
  \end{tabular}
  \caption{Multiplication table of discrete symmetries. $\mathbb{1}$ denotes the identity operation, $I$ denotes inversion of space.}\label{SymmetryGroupTable}
\end{table}
 
 Because inversion of space can be written as a composition of the up-down mirror symmetry and the rotation $R_n$, $I=R_n M_n$, inversion of space and up-down mirror symmetry are preserved and broken simultaneously for a surface with rotation symmetry in the plane. 
  
Under these symmetries, the stress, torque, curvature, Levi-Civita tensor, as well as vectors and pseudo vectors are modified. We list in Table \ref{SignatureSymmetryTable} the signatures of transformations of components of these tensors under the symmetries introduced above. Additional transformations arise from the combinations $M_n R_n$, $M_t R_n$, and $R_t R_n$, which are not relevant to discuss symmetry properties of our equations.
  
 \begin{table}[h!!]
\centering
\begin{tabular}{ccccc}
Symmetry & $M_n$ & $M_t$ & $R_t$  &$R_n$\\\hline
$t_{ij}$ & 1 & 1 & 1 &1\\
$t_{n}^i$ & -1 & 1 & -1  &-1 \\
$m_{ij}$  & -1 & -1 & 1 &1 \\
$m_{n}^i$  & 1 & -1 & -1  &-1 \\
$C_{ij}$ & -1 & 1 & -1&1 \\
$\epsilon_{ij}$ & 1 & -1 & -1  &1 \\
$v^i$ & 1 & 1 & 1 &-1\\
$v_n$ & -1 & 1 & -1 &1 \\
$\nabla_i$ & 1 & 1 & 1  &-1\\
$v_{ij}$ & 1 & 1 & 1  &1\\
$\omega_{i}$ & -1 & -1 & 1  &-1\\
$\omega_{n}$ & 1 & -1 & -1  &1
\end{tabular} 
\caption{Signature of symmetries on vector and tensor fields of the surface.}\label{SignatureSymmetryTable}
\end{table}

The force and torque balance equations \ref{ForceBalanceTangential}-\ref{TorqueBalanceNormal} are invariant under these transformations. One can further verify that transformations under $R_n$ preserve all the constitutive equations \ref{ConstitutiveEquationtij}-\ref{ConstitutiveEquationj}.
\section{Equilibrium tension and moment tensors, external force and torque surface densities for a fluid membrane}
\label{PassiveMembrane}

Using the expression of the virtual work given in Eq. \ref{VirtualWork2}, we obtain in this appendix the equilibrium tension tensor and moment tensor for a fluid membrane, first for the generic case, and then  for the specific case of an Helfrich membrane. We then obtain the external force and torque surface densities induced by an external potential acting on the surface.

\subsection{Tension and moment tensors for a generic equilibrium fluid membrane}

We start here from a fluid membrane with a free energy density given by Eq. \ref{FluidMembraneFreeEnergyDensity}, such that the free energy for a region of surface $\mathcal{S}$ is given by $
F=\int_{\mathcal{S}} dS f_0$
with $df_0=\mu^{\alpha} d c^{\alpha}+K^{i}{}_{j} dC_{i}{}^{j}-sd T$.
We now calculate the change of free energy following a change of shape of the surface. A change of the surface metric results in a dilution of the concentrations, according to $\delta c^{\alpha}/c^{\alpha}=-\delta(\sqrt{g})/\sqrt{g}$. As a result and using Eq. \ref{VariationSqrtDeterminantMetric}, the free energy differential following a shape change reads
\begin{equation}
\delta F=\int_{\mathcal{S}} dS \left[(f_0 - \mu^{\alpha} c^{\alpha} )g^{ij} \frac{\delta g_{ij}}{2} +K^{i}{}_{j} \delta C_{i}{}^{j}\right]
\end{equation}

At equilibrium, inertial forces vanish and we have $\delta W=\delta F$ for infinitesimal deformations. Using Eq. \ref{VirtualWork2}, one can then identify the equilibrium tensors $\bar{t}_e^{ij}$, $\bar{m}_e^{ij}$, $m_{e,n}^i$ given in Eqs. \ref{FluidEquilibriumInPlaneStressTensor}-\ref{FluidEquilibriumNormalTorqueTensor}.

\subsection{Tension and moment tensors for a Helfrich membrane}
The Helfrich free energy functional for a region of surface $\mathcal{S}$ of a fluid membrane reads:
\begin{eqnarray}
F&=&\int_\mathcal{S} dS\bigg[ \gamma+\kappa \left(\left(C_i{}^i\right)^2-4C_0C_i{}^i\right)+2\kappa_g \det(C_i{}^j)\bigg],
\end{eqnarray}
where $\gamma$ is the surface tension, $\kappa$ is the bending rigidity, $C_0$ is the spontaneous curvature and $\kappa_g$ the gaussian bending modulus. Using relation Eq. \ref{VirtualWork2} and the relation $\delta W=\delta F$ for infinitesimal deformations, one finds for the in-plane stress tensor and bending moment tensor:
\begin{eqnarray}
\bar{t}_e^{ij}&=&\big(\gamma+\kappa \left((C_k{}^k)^2-4 C_0 C_k{}^k\right) +2\kappa_g \det(C_k{}^l)\big)g^{ij}, \\
\bar{m}_{e}^{ij}&=&2(\kappa+\kappa_g) C_k{}^k g^{ij}-4\kappa C_0 g^{ij}- 2\kappa_g C^{ij},
\end{eqnarray}
where we have used $\det(C_i{}^j) =\frac{1}{2}\left[ (C_k{}^k)^2- C_{ik} C^{ki}\right]$ by taking the trace of Eq. \ref{DesernoGaussEquation}, and Eq. \ref{VariationSqrtDeterminantMetric}. Note that the stress tensor $t_{ij}$ and bending moment tensor $m_{ij}$ are then given by
\begin{eqnarray}
\label{tijHelfrichExpression}
t_e^{ij}&=&\gamma  g^{ij} -2\kappa  (C_k{}^k-2C_0) C^{ij}+\kappa  C_k{}^k ( C_k{}^k-4C_0) g^{ij},\\
m_e^{ij}&=&2(\kappa+\kappa_g) C_k{}^k \epsilon^{ij}-4\kappa C_0 \epsilon^{ij}- 2\kappa_g C^{ik} \epsilon^{kj},\label{mijHelfrichExpression}
\end{eqnarray}
where we have used the identity \ref{DesernoGaussEquation}. The stress tensor $t_{ij}$ therefore does not depend on the gaussian bending modulus $\kappa_g$.
Furthermore, from the force balance equation \ref{TorqueBalanceTangential} and because of Eq \ref{DivergenceCurvatureTensor}, the normal shear stress $\mathbf{t}_n$ is given in the absence of external torque by
\begin{eqnarray}
\label{tnHelfrichExpression}
 t_{n}^j&=&\nabla_i \bar{m}_e^{ij}=2\kappa \nabla^j C_k{}^k,
\end{eqnarray}
which also does not depend on the Gaussian bending modulus. Eqs. \ref{tijHelfrichExpression} and \ref{tnHelfrichExpression} are in accordance with Ref. \cite{capovilla2002stresses}, with an opposite sign convention for the force density $\mathbf{t}$. 

Note that while $m^{ij}_e$ depend on $\kappa_g$, terms proportional to $\kappa_g$ cancel when using force balance equations \ref{ForceBalanceEquation} and \ref{TorqueBalanceEquation}. Therefore, in accordance with the Gauss-Bonnet theorem, the Gaussian bending modulus only enters boundary conditions when solving the force balance equations to find the the surface shape.

\subsection{External force and torque density for an equilibrium fluid membrane}

If molecules in the surface are subjected to an external potential $U=c^{\alpha} U^{\alpha}(s^1,s^2,\mathbf{n})$, where $U^{\alpha}$ acts on component $\alpha$, the variation of this external potential induced by a deformation of the surface element $\mathcal{S}$ reads
\begin{eqnarray}
\delta U&=&\int_{\mathcal{S}} dS c^{\alpha} \left[(\partial_i U^{\alpha}) \delta X^i + (\partial U^{\alpha}/\partial \mathbf{n})\cdot \delta \mathbf{n}\right],\nonumber\\
&=&\int_{\mathcal{S}} dS c^{\alpha} \big[(\partial_i U^{\alpha}) \delta X^i + \frac{1}{2}((\partial U^{\alpha}/\partial \mathbf{n})\cdot \mathbf{e}^i)\epsilon_{ik} \mathbf{e}^k \cdot (\boldsymbol{\nabla}\times \delta \mathbf{X})\big].
\end{eqnarray}
where we have used the identities \ref{VariationNormalVectorFirstExpression} and \ref{TangentialCurlInfinitesimalDisplacement}. The contribution of external forces and torques to the virtual work given in Eq. \ref{VirtualWorkDefinition} on the other hand reads, ignoring here inertial terms,
\begin{eqnarray}
\delta W_{ext}=\int_\mathcal{S} dS \mathbf{f}^{\rm{ext}} \cdot\delta\mathbf{X}
+\frac{1}{2} \boldsymbol{\Gamma}^{\rm{ext}} \cdot(\boldsymbol{\nabla}\times \delta\mathbf{X})).
\end{eqnarray}
Using $\delta W_{ext}=-\delta U$, one then obtains the external surface force density and external surface torque density:
\begin{eqnarray}
\mathbf{f}^{\rm{ext}}&=&-c^{\alpha}\partial_i U^{\alpha} \mathbf{e}^i\label{ExternalForceDensityEquilibriumFluidMembrane},\\
\boldsymbol{\Gamma}^{\rm{ext}}&=&-c^{\alpha}((\partial U^{\alpha}/\partial \mathbf{n})\cdot \mathbf{e}^j)\epsilon_{ji} \mathbf{e}^i.\label{ExternalTorqueDensityEquilibriumFluidMembrane}
\end{eqnarray}

\section{Entropy production rate for a fluid surface}
\label{AppendixEntropyProductionRate}
We derive here the entropy production rate for a region of a fluid surface $\mathcal{S}$ enclosed by a fixed contour $\mathcal{C}$. The time derivative of the free energy of the surface $F$ is
\begin{align}
\frac{dF}{dt}&=\int_\mathcal{S} dS \bigg[\frac{1}{2}\partial_t (\rho v_i v^i+\rho (v_n)^2)+K^{i}{}_{j} \partial_t C_{i}{}^{j}+\mu^{\alpha} \partial_t c^{\alpha}\bigg]+\int_\mathcal{S} dS v_n C_i{}^i \left[\frac{1}{2}\rho v^2+f_0\right],
\end{align}
Using the following relation obtained from Eq. \ref{TimeDerivativeMetricEulerian} $
\partial_t v_i = \partial_t (g_{ij} v^j)=g_{ij} \partial_t v^j + 2 v_n C_{ij} v^j$, as well as
the mass balance equation \ref{MassBalanceEquation}, one obtains
\begin{eqnarray}
\frac{1}{2}\partial_t (\rho v_i v^i+\rho (v_n)^2)+\frac{1}{2}\rho v_n C_i{}^i v^2=\rho a^i  v_i + \rho  a_n v_n -\frac{1}{2}\nabla_i(\rho v^2 v^i) +\frac{1}{2} J_n^{\rho} v^2,
\end{eqnarray}
where we have used the expression of the acceleration $\mathbf{a}$ obtained in Eq. \ref{DetailedEulerianExpressionAcceleration}.
Using then the force balance equation \ref{ForceBalanceEquation} and the concentration balance equation \ref{ConcentrationBalanceEquation}, we find
\begin{align}
\frac{dF}{dt}&=\int_\mathcal{S} dS \bigg[-\frac{1}{2} \nabla_i(\rho v^i v^2 ) +\frac{1}{2}J_n^{\rho}\mathbf{v}^2+ \nabla_j t^{ji} v_i+t_n^j  C_j{}^i v_i + f^{\rm{ext},i} v_i+  \nabla_i t^i_n v_n- t^{ij}C_{ij} v_n+f^{\rm{ext}}_{n} v_n \nonumber\\
&+K^{i}{}_{j} \partial_t C_{i}{}^{j}+(f_0- \mu^{\alpha}c^{\alpha}) C_i{}^i v_n-\mu^{\alpha} \nabla_i(c^{\alpha} v^i +j^{\alpha,i}) +\mu^{\alpha} (J_n^{\alpha}+r^{\alpha})\bigg].
\end{align}
Using the divergence theorem \ref{DivergenceTheorem}, this can be rewritten
\begin{align}
\frac{dF}{dt}&=\int_\mathcal{S} dS \bigg[-  t^{ji} \nabla_j v_i+t_n^j C_j{}^i v_i -t^i_n\partial_i v_n -t^{ij}C_{ij}v_n +\mathbf{f}^{\rm{ext}}\cdot\mathbf{v}+ (f^0 - \mu^{\alpha}c^{\alpha}) C_i{}^i v_n+K^{i}{}_{j} \partial_t C_{i}{}^{j}\nonumber\\
&+ (\partial_i \mu^{\alpha}) (c^{\alpha} v^i +j^{\alpha,i}) +\mu^{\alpha} \left(J_n^{\alpha}+r^{\alpha}\right)+\frac{1}{2}J_n^{\rho}v^2\bigg]+\oint_\mathcal{C} d\nu_i\left[- \frac{1}{2}\rho v^2 v^i+ t^{ij} v_j + t_n^i v_n-\mu^{\alpha} (c^{\alpha} v^i+j^{\alpha,i})\right].
\end{align}
Using the Gibbs-Duhem equality \ref{GibbsDuhemTranslation} and the balance of fluxes \ref{BalanceNormalFluxes} and \ref{BalanceChemicalReactions},
\begin{align}
\frac{dF}{dt}&=\int_\mathcal{S} dS \bigg[ -   t^{ij}\nabla_i v_j -t^{ij}C_{ij} v_n+ t_n^i (C_i{}^j v_j-\partial_i v_n)  +\mathbf{f}^{\rm{ext}}\cdot\mathbf{v}+ (f_0 - \mu^{\alpha}c^{\alpha})  C_i{}^i v_n+K^{i}{}_{j} \partial_t C_{i}{}^{j}\nonumber\\
&+ v^i K_{jk}\nabla_i C^{jk}-(\partial_i (f_0 -\mu^{\alpha} c^{\alpha}))v^i + (\partial_i \mu^{\alpha}) j^{\alpha,i} +(\mu^{\alpha}+\frac{1}{2} m^{\alpha} v^2) J_n^{\alpha}+\mu^{\alpha} r^{\alpha}\bigg]\nonumber\\
&+\oint_\mathcal{C} ds_i\left[ -\frac{1}{2}\rho v^2 v^i+ \mathbf{t}^i \cdot \mathbf{v}-\mu^{\alpha} (c^{\alpha} v^i+j^{\alpha,i})\right].
\end{align}
Reorganizing, performing an integration by part, introducing the convected derivative of the curvature tensor, 
\begin{equation}
\frac{d C_{i}{}^{j}}{dt}=\partial_t C_{i}{}^{j}+ v^k \nabla_k C_{i}{}^{j},
\end{equation}
and using the total chemical potential $\mu_{tot}^{\alpha}=\mu^{\alpha}+\frac{1}{2} m^{\alpha} v^2$, one finds:
\begin{align}
\frac{d F}{dt}&=\int_\mathcal{S} dS \bigg[ - \left[ t^{ij}-(f_0-\mu^{\alpha}c^{\alpha})g^{ij}\right]\nabla_i v_j -t^{ij} C_{ij}v_n +t^i_n(C_i{}^j v_j -\partial_i v_n) \nonumber\\
&+\mathbf{f}^{\rm{ext}}\cdot\mathbf{v}+ (f_0- \mu^{\alpha}c^{\alpha}) C_i{}^i v_n +K^{i}{}_{j}\frac{d C_{i}{}^{j}}{dt}+ (\partial_i \mu^{\alpha}) j^{\alpha,i} +\mu^{\alpha}_{tot} J_n^{\alpha}+\mu^{\alpha} r^{\alpha}\bigg]\nonumber\\
&+\oint_\mathcal{C} ds_i \left[ -\frac{1}{2}\rho v^2 v^i+ \mathbf{t}^i \cdot \mathbf{v}-\mu^{\alpha} j^{\alpha,i} -f_0 v^i\right].
\end{align}
Using the torque balance equation \ref{TorqueBalanceEquation}, splitting the tension tensor $t^{ij}$ into a symmetric and an antisymmetric part, and introducing the equilibrium tension tensor $\bar{t}^{ij}_e=(f_0-\mu^{\alpha} c^{\alpha} )g^{ij}$, we obtain
\begin{align*}
\frac{dF}{dt}
&=\int_\mathcal{S} dS \bigg[- (t_s^{ij}-\bar{t}^{ij}_e) v_{ij} -\frac{1}{2} t^{kl}\epsilon_{kl}\epsilon^{ij} \nabla_i v_j+(\nabla_k \bar{m}^{ki}-C_l{}^k m_n^l \epsilon^{ki} -\Gamma^{\rm{ext},k} \epsilon_k{}^i )(C_i{}^j v_j-\partial_i v_n)\\
&+K^{i}{}_{j} \frac{d C_{i}{}^{j}}{dt}+\mathbf{f}^{\rm{ext}}\cdot\mathbf{v}+ (\partial_i \mu^{\alpha}) j^{\alpha,i}+\mu^{\alpha}_{tot} J_n^{\alpha}+\mu^{\alpha} r^{\alpha}\bigg]+\oint_\mathcal{C} dl\nu_i\left[ -f v^i+ \mathbf{t}^i \cdot \mathbf{v}-\mu^{\alpha} j^{\alpha,i} \right],
\end{align*}
where we have introduced the symmetric velocity gradient $v_{ij}$ defined in Eq. \ref{DefinitionSymmetricVelocityGradient} and used the symmetry of the curvature tensor, $\epsilon_{ij} C^{ij}=0$, and the symmetry of the tensor $KC$ implied by rotational invariance (Eq. \ref{GibbsDuhemRotation}). Using the torque balance equation \ref{TorqueBalanceNormal}, performing an integration by part, and using the definition of the vorticity of the flow (Eq. \ref{DefinitionFlowRotational}),
\begin{align}
\frac{dF}{dt}
&=\int_\mathcal{S} dS \bigg[-(t_s^{ij} -\bar{t}^{ij}_e)v_{ij}+(\nabla_i m^i_n-C_{ij} \bar{m}^{ik}\epsilon_k{}^j+\Gamma^{\rm{ext}}_n)\omega_n-\bar{m}^{ki}\nabla_k(C_i{}^j v_j -\partial_i v_n)  \nonumber \\
&+m_n^j C_j{}^i  \omega_i+K^{i}{}_{j} \frac{d C_{i}{}^{j}}{dt}+\mathbf{f}^{\rm{ext}}\cdot\mathbf{v} +\Gamma^{\rm{ext},i}  \omega_i+ (\partial_i \mu^{\alpha}) j^{\alpha,i}+\mu^{\alpha}_{tot} J_n^{\alpha}+\mu^{\alpha} r^{\alpha}\bigg]\nonumber\\
&+\oint_\mathcal{C} dl\nu_i \left[-f v^i+\mathbf{t}^i \cdot \mathbf{v}+\bar{m}^{ik}(C_k{}^j v_j-\partial_k v_n) -\mu^{\alpha} j^{\alpha,i} \right]. 
\end{align}
Rearranging and performing an integration by part,
\begin{align}
\frac{dF}{dt}
&=\int_\mathcal{S} dS \bigg[-(t_s^{ij}-\bar{t}^{ij}_e+C_k{}^j \bar{m}^{ki})v_{ij}-\bar{m}^{i}{}_{j} \left[ -\nabla_i (\partial^j v_n)- C_{ik} C^{jk}  v_n+v_k\nabla_i C^{jk}  + (\nabla_i v^k) C^k{}_j -(\nabla_k v_j) C_{ik} \right]\nonumber\\
& - m_n^i \partial_i \omega_n +m_n^i C_i{}^j \omega_j+K^{i}{}_{j} \frac{d C_{i}{}^{j}}{dt}+\mathbf{f}^{\rm{ext}}\cdot\mathbf{v}+\boldsymbol{\Gamma}^{\rm{ext}}\cdot\boldsymbol{\omega}+ (\partial_i \mu^{\alpha}) j^{\alpha,i}+\mu^{\alpha}_{tot}J_n^{\alpha}+\mu^{\alpha} r^{\alpha}\bigg]\nonumber\\
&+\oint_\mathcal{C} dl\nu_i \left[-f v^i+ \mathbf{t}^i \cdot \mathbf{v} + m^{ij}\omega_j+m_n^i \omega_n-\mu^{\alpha} j^{\alpha,i} \right]\label{TempEquationdFdt} .
\end{align}
In Eq. \ref{TempEquationdFdt}, the term in factor of $\bar{m}^{i}{}_{j}$ corresponds to the Lagrangian convected derivative of the curvature tensor, $\bar{D} C_{i}{}^{j}/\bar{D}t$, defined in Eq. \ref{LagrangianDerivativeCurvatureTensor}:
\begin{eqnarray}
\frac{\bar{D} C_{i}{}^{j}}{\bar{D}t}&=&\frac{d C_i{}^j}{dt}+(\nabla_i v^k) C_k{}^j -(\nabla_k v^j) C_{ik},
\end{eqnarray}
where we have used the Mainardi-Coddazzi equation \ref{MainardiCoddazi}.
We define the bending rate tensor as the symmetric part of this tensor:
\begin{eqnarray}
\frac{D C_{ij}}{Dt}&=&\frac{1}{2}\left(g_{jk} \frac{\bar{D}C_i{}^k}{\bar{D}t}+ g_{ik} \frac{\bar{D}C_j{}^k}{\bar{D}t}\right),
\end{eqnarray}
whose explicit expression is given in Eq. \ref{DefinitionCorotationalCurvatureTensor}.
In addition, one can verify that $K^i{}_j \bar{D} C_i{}^j/\bar{D}t=K^i{}_j  d C_i{}^j/dt$; indeed
\begin{eqnarray}
K^i{}_j\left[(\nabla_i v_k) C^{kj}-(\nabla_k v_j) C^{ki}\right]=\nabla_i v_j \left[K^{ik}C^j{}_k-K^{kj}C^i{}_k\right]=0,
\end{eqnarray}
as a result of the invariance by rotation, Eq. \ref{GibbsDuhemRotation}, and the symmetry of $K^{ij}$. 
Using these relations and the symmetry of the tensor $K^{ij}$, we then find the expression for the rate of change of free energy:
\begin{eqnarray}
\frac{dF}{dt}
&=&\int_\mathcal{S} dS \bigg[-(\overline{t}^{ij}-\overline{t}^{ij}_e + \frac{1}{2}\epsilon_{ln} \bar{m}^{ln} (\epsilon^{ik} C_{k}{}^{j} +\epsilon^{jk} C_{k}{}^{i} ))v_{ij}- (\bar{m}^{ij}-K^{ij})\frac{D C_{ij}}{Dt} -m_n^i \left[  \partial_i \omega_n -C_i{}^j \omega_j \right]  \nonumber\\
&&+ \mathbf{f}^{\rm{ext}}\cdot\mathbf{v}+ \boldsymbol{\Gamma}^{\rm{ext}}\cdot \boldsymbol{\omega}+ (\partial_i \mu^{\alpha}) j^{\alpha i}+\mu^{\alpha}_{tot} J_n^{\alpha}+\mu^{\alpha} r^{\alpha}\bigg] +\oint_\mathcal{C} dl\nu_i \left[-f v^i+  \mathbf{t}^i \cdot \mathbf{v} +\mathbf{m}^{i} \cdot \boldsymbol{\omega}-\mu^{\alpha} j^{\alpha i}\right]. 
\end{eqnarray}
In Eq. \ref{RateDissipationFreeEnergy}, we have not included the contribution $\epsilon_{ln}\bar{m}^{ln}$ of the antisymmetric part of $\bar{m}^{ij}$. The antisymmetric part of $\bar{m}^{ij}$ is related to the trace of $m^{ij}$ through the relation $m_k{}^k=-\bar{m}_{ij}\epsilon^{ij}$. 
The transformation invariance in Eqs. \ref{GaugeTransformationtij}-\ref{GaugeTransformationmij} implies that the contribution of the antisymmetric part of the bending moment tensor $\bar{m}_{ij}$ to the force balance equation can be absorbed in a redefinition of the stress tensor.

\section{Stability of an homogeneous flat active surface}
\label{LinearStability}
We discuss here the stability of a homogeneous flat active surface, in the absence of external forces and torques. 
Perturbations of the flat shape are described in the Monge gauge by the height $h(x,y)$ such that the surface position is given by $\mathbf{X}(x,y)=x \mathbf{u}_x+y\mathbf{u}_y+h(x,y)\mathbf{u}_z$. Calculations are performed for a weakly bent surface, $|\partial_i h|\ll 1$, at linear order in the height $h$ and velocity $\mathbf{v}$.
In this limit, covariant and contravariant indices can be used indifferently, and
\begin{eqnarray}
g_{ij}&\simeq& \delta_{ij}\\
C_{ij}&\simeq&-\partial_i \partial_j h\\
v_{ij}&\simeq&\frac{1}{2}(\partial_i v_j+\partial_j v_i)\\
\omega_n&\simeq&\frac{1}{2}\epsilon_{ij} \partial_i v_j\\
\omega_i&\simeq&\epsilon_{ij} \partial_j v_n.
\end{eqnarray}
\subsubsection{Fluid surface}
For a fluid surface, the tensions and torque tensors are given by
\begin{eqnarray}
\overline{t}_{ij}&=&\eta (\partial_i v_j+\partial_j v_i) + (\eta_b-\eta) \partial_k v_k \delta_{ij}-2\bar{\eta} \partial_i\partial_j \partial_t h-(\bar{\eta}_b-\bar{\eta}) (\partial_t \Delta h )\delta_{ij}\nonumber\\
&&+ \left(\gamma+\zeta\Delta\mu-\left(-4\kappa C_0 +(\zeta'-\tilde{\zeta}) \Delta\mu\right)  \Delta h\right) \delta_{ij} - 2\tilde{\zeta} \Delta \mu \partial_i \partial_j h \label{ConstitutiveEquationtij4}\\
\bar{m}_{ij}&=&-2\eta_c \partial_i\partial_j \partial_t h-(\eta_{cb}-\eta_c) (\partial_t \Delta h) \delta_{ij}+\bar{\eta} (\partial_i v_j+\partial_j v_i) + (\bar{\eta}_b-\bar{\eta}) \partial_k v_k \delta_{ij}+\nonumber\\
&&+\left(\zeta_c \Delta\mu-4\kappa C_0-\left(2\kappa+2\kappa_g+(\zeta_c' -\tilde{\zeta}_c) \Delta\mu\right) \Delta h \right) \delta_{ij}  -2\left[-\kappa_g+ \tilde{\zeta_c}\Delta \mu \right] \partial_i \partial_j h \label{ConstitutiveEquationmij4}\\
m_{n,i}&=&\frac{\lambda}{2} \partial_i \epsilon_{kl} \partial_k v_l.
\end{eqnarray}
where we have used the Laplacian operator $\Delta=\partial_x^2+\partial_y^2$. The force and torque balance equations then yield, 
neglecting inertial terms at low Reynolds number
\begin{eqnarray}
\eta \Delta v_j+\eta_b\partial_j \partial_k v_k-(\bar{\eta}_b+\bar{\eta}) \partial_j \partial_t  \Delta h -(\zeta'+\tilde{\zeta}-\zeta_c) \Delta\mu\partial_j \Delta h=-\frac{1}{2} \epsilon_{ij} \partial_i( \epsilon_{kl}t_{kl}) \\
\partial_i t_{n,i}=-\Delta h\left[\gamma+\zeta\Delta\mu\right]\\
t_{n,j} =-(\eta_{cb}+\eta_c)  \partial_j\partial_t \Delta h +\bar{\eta} \Delta v_j +\bar{\eta}_b\partial_j \partial_i v_i-
\left[2\kappa+(\zeta_c'  +\tilde{\zeta_c})\right]  \partial_j \Delta h\\
\epsilon_{ij} t_{ij}=-\frac{\lambda}{2}  \Delta(\epsilon_{kl} \partial_k v_l) .
\end{eqnarray}
We then obtain the shape equation
\begin{eqnarray}
\left[\eta_{cb}+\eta_c-\frac{(\bar{\eta}+\bar{\eta}_b)^2}{\eta+\eta_b} \right] \partial_t \Delta \Delta h &=&-
\left(2\kappa+\left(\zeta_c'  + \tilde{\zeta_c} +\frac{\bar{\eta}+\bar{\eta}_b}{\eta+\eta_b} (\zeta_c-\zeta'-\tilde{\zeta})\right)\Delta\mu \right)  \Delta \Delta h\nonumber\\
&&\hspace{4cm}+\left(\gamma+\zeta\Delta\mu\right)\Delta h.
\end{eqnarray}
\subsubsection{Elastic surface}
For an elastic surface, the tensions and torque tensors are given in the limit of small displacements by
\begin{eqnarray}
\overline{t}_{ij}&=&E_1 \frac{1}{2}(\partial_i u_j+\partial_j u_i) +E_2 \partial_k u_k \delta_{ij} - G_1 \partial_i \partial_j h - G_2 \Delta h \delta_{ij} + \Delta \mu (\zeta \delta_{ij} -( \zeta'-\tilde{\zeta}) \Delta h \delta_{ij}-2 \tilde{\zeta} \partial_i \partial_j h) \\
\overline{m}_{ij}&=&-F_1 \partial_i \partial_j h -F_2 \Delta h \delta_{ij} +K_1 \frac{1}{2}(\partial_i u_j+\partial_j u_i) + K_2 \partial_k u_k \delta_{ij}+\Delta \mu(\zeta_c \delta_{ij}-(\zeta_c'-\tilde{\zeta}_c) \Delta h \delta_{ij} - 2\tilde{\zeta}_c \partial_i \partial_j h) \\
m_{n,i}&=&H_1 \frac{1}{2} \epsilon_{jk} \partial_i \partial_j u_k\quad.
\end{eqnarray}
A calculation similar to the fluid case then yields the equation for the surface height
\begin{equation}
\xi \frac{dh}{dt}=\zeta\Delta\mu \Delta h- \left[F_1+F_2+\zeta_c'+\tilde{\zeta}_c -\frac{K_1+K_2}{E_1+E_2}\left(G_1+G_2+(\zeta'+\tilde{\zeta}-\zeta_c)\Delta\mu\right)\right]\Delta\Delta h \quad,
\end{equation}
where we have introduced an effective external friction force $\mathbf{f}^{\rm{ext}}=-\xi\mathbf{v}\cdot\mathbf{n}$.
\bibliographystyle{unsrt}
\bibliography{ActiveSurfaces}
\end{document}